\documentclass[final,3p,times,twocolumn]{elsarticle}

\usepackage{lineno}
\modulolinenumbers[5]

\usepackage{graphicx}
\usepackage[utf8]{inputenc}
\usepackage{textcomp}

\usepackage{booktabs}
\usepackage{rotating}
\usepackage{multirow}
\usepackage{subfigure}

\usepackage{amsmath,amssymb}
\usepackage{gensymb}	

\usepackage{url,hyperref}

\usepackage{color}
\usepackage[usenames,dvipsnames,svgnames,table]{xcolor}

\journal{Nuclear Instruments and Methods in Physics Research Section B}

\begin{document}

\begin{frontmatter}

\title{Cu self-sputtering MD simulations for 0.1--5 keV ions at elevated temperatures}




\author[Tartu]{Tarvo Metspalu}
\author[HIP]{Ville Jansson\corref{mycorrespondingauthor}}
\ead{ville.b.c.jansson@gmail.com}
\author[Tartu,HIP]{Vahur Zadin}
\author[HIP]{Konstantin Avchaciov}
\author[HIP]{Kai Nordlund}
\author[Tartu]{Alvo Aabloo}
\author[HIP,Moscow]{Flyura Djurabekova}
\ead{flyura.djurabekova@helsinki.fi}

\cortext[mycorrespondingauthor]{Corresponding author}


\address[Tartu]{Intelligent Materials and Systems Lab, Institute of Technology, Tartu University, Nooruse 1, 50411 Tartu, Estonia}
\address[HIP]{Helsinki Institute of Physics and Department of Physics, P.O. Box 43 (Pehr Kalms gata 2), FI-00014 University of Helsinki, Finland}
\address[Moscow]{National Research Nuclear University MEPhI, Kashirskoye sh. 31, 115409 Moscow, Russia}

\begin{abstract}
Self-sputtering of copper under high electric fields is considered to contribute to plasma buildup during a vacuum breakdown event frequently observed near metal surfaces, even in ultra high vacuum condition in different electric devices. In this study, by means of molecular dynamics simulations, we analyze the effect of surface temperature and morphology on the yield of self-sputtering of copper with ion energies of 0.1--5 keV. We analyze all three low-index surfaces of Cu, \{100\}, \{110\} and \{111\}, held at different temperatures, 300 K, 500 K and 1200 K. The surface roughness relief is studied by either varying the angle of incidence on flat surfaces, or by using arbitrary roughened surfaces, which result in a more natural distribution of surface relief variations. Our simulations provide detailed characterization of copper self-sputtering with respect to different material temperatures, crystallographic orientations, surface roughness, energies, and angles of ion incidence.
\end{abstract}


\end{frontmatter}


\section{Introduction}

Particle accelerators, free electron lasers, fusion reactors and other devices using high electric fields in ultra high vacuum, often experience spontaneous plasma discharges happening on the material surface, known as vacuum arcing or electrical breakdowns. Surface electric fields in these devices can reach up to 300 MV/m as, e.g., in the accelerating structures of the Compact Linear Collider (CLIC) with the target energy of electron-positron collisions of 0.5--5 TeV \cite{clic2016updated}.  It is strongly believed that the breakdowns are preceded by strong field emission \cite{Kil04b_fd,Mura11_fd}, however, it is still a puzzle how the plasma densities build up in the ultrahigh vacuum condition. Even though the very initial atoms may be found at random above the surface, these are not sufficient to develop a full self-sustainable plasma, detected as electrical breakdowns disturbing the operation of the machine \cite{Ship12_fd,Shipman2014ExperimentalCLIC, Tim11b}. It is believed that the plasma is fueled by atoms sputtered from the surface by ions forming above the surface due to interaction with electrons \cite{Tim11b}. The main component of such a plasma are the ions of the surface material. Thus the self-sputtering may play a crucial role for understanding of the plasma buildup process.

It has been proposed that the increase of the surface temperature through field emission and plasma arc heating may intensify the amount
of neutral Cu atoms emerging from Cu self-bombardment \cite{Tim11b}. In other words,
the temperature might be increasing the sputtering yield --- the number of atoms kicked out of the surface per incident ion accelerated by the plasma sheath potential \cite{Sigmund2012RecollectionsSputtering}. Experimentally, thermally enhanced sputtering has been observed in linear plasma devices \cite{Temmerman12Hebubbles}.

It is well known that sputtering yields depend on many irradiation parameters, such as incident angle, energy, and mass of the incoming ion, as well as the mass and surface binding energy of the target atoms \cite{Zykova2010MassProperties}. In addition, for a crystalline target such as Cu, the crystallographic orientation of the
surface with respect to the direction of the projectile may play an important role \cite{Sigmund1993FundamentalMolecules}.
Moreover, surface roughness may also affect the sputtering yield. For example, a study by Makeev and Barabási \cite{Makeev2004EffectSurfaces}, found an up to 100 \% increase of the sputtering yield, compared to flat surface values.

Currently, no fundamental equation, which would be able to predict the effect of all the aforementioned variables on the sputtering yield, exist --- although several phenomenological expressions developed to fit experimental observations were suggested throughout the last few decades \cite{yamamura1983theoretical,yamamura1983angular,Yamamura1996EnergyIncidence,Eckstein2003NewYield,Garcia-Rosales1995RevisedData,Grais2013AAngle}.
These models work best for the elements, which are neither too heavy (as, for instance, Au) nor too light (as, e.g. H, He), with not too low energy of incident ions, and the target surface being free of any contaminants. One of the major disadvantages of the existing semi-empirical models is the lack of a description of the temperature dependence of sputtering yields, although this dependence sometimes can show intriguing patterns \cite{Lee1993Y_Ag(T)}. Comparing the available fits to experimental data for Cu self-sputtering in Refs.\,\cite{Yamamura1996EnergyIncidence,Eckstein2003NewYield,matsunami1984energy}, the formula by Yamamura \cite{yamamura1983theoretical, matsunami1984energy} shows the best agreement.

At elevated temperatures, the thermal vibration energies also increases, leading towards the decrease of the effective binding energy on the surface. However, as the experiments suggest, this explanation is not sufficient to explain the temperature dependence of sputtering yield \cite{Behrisch2007SputteringEnergies,Temmerman12Hebubbles}.

It has also been hypothesized that at elevated temperatures, the kinetic
energy of the projectile is more easily dissipated, i.e. there are less focused
few-body collisions and the sputtering yield drops \cite{Carlston1965EffectYields}.
On the other hand, it was also shown that in thermal spikes, which can be expected in metal irradiated by ions with energies higher than 2 keV, ambient thermal effects are negligible as the temperature of atoms in the spikes is very high \cite{Ave98,Gad95}. Normally, elevated temperatures may cause annealing of radiation defects, especially in metals, which may potentially decrease the sputtering yield, however, this process is usually neglected as small defects created in the cascades move relatively fast already at room temperature \cite{sigmund1984temperature,Sigmund1984Temperature-dependentInsulators}.

A computer simulation study indicates up to $\sim$10 times increase of the sputtering yield as compared to the value at room temperature \cite{Insepov2010AtomisticLinacs,Staudt2000OnYields}.
Additionally, an exponential increase of the sputtering yield due to thermal evaporation has been observed near melting temperatures, although in this regime ballistic and thermal effects may overlap  \cite{Sigmund1993FundamentalMolecules,Tem12}.

Experimentally it has been reported that for Cu and Al --- both face centered cubic
(FCC) metals --- the sputtering yield either stays rather constant \cite{Bohdansky1987_Cu-Li(T)} or experiences a small-scale deviation with
an increase of the target surface temperature (up to 20--30 \%) \cite{Ph.Elich1971Copper600K}. Furthermore, no difference in the sputtering yield with increasing target temperature was observed for the \{100\} and \{110\} surfaces bombarded at the normal incidence, while a minor decrease in the sputtering yield with increasing temperature was observed for the normal bombardment of the \{111\} surface. This clearly indicates an anisotropy in temperature dependencies for different surface orientations \cite{Carlston1965EffectYields}.

The aim of this study is to test the hypothesis that the
temperature has a dramatic effect on the primary sputtering process during
the self-bombardment process of
Cu, the main material for accelerating structures of CLIC, and thereby produces more unbound atoms
upon which an electric breakdown can occur. We focus in this paper on the effects
induced directly by collision cascades, i.e. we do not in this paper examine the possibility that
thermal evaporation may enhance atom desorption between the actual collision
cascades \cite{Sch06b} or during overlapping cascades \cite{Tim09a}. However,
the possible role of these effects will be touched on in the discussion section.

In Sec.~\ref{sec:materials_methods}, we describe the molecular dynamics simulation setup and the interatomic potentials used in the current study. In Sec.~\ref{sec:angular}, we investigate the angular dependencies of the sputtering yields by considering a wide range of different incident ion angles at different temperatures and different surface crystallographic orientations. In Sec.~\ref{sec:channeling}, we study the channeling effects for different crystallographic orientations. The dependence of the target temperature is investigated in Sec.~\ref{sec:temperature} and in Sec.~\ref{sec:roughness} we consider the effect of surface roughness on sputtering yield. The results are discussed and the paper is concluded in Sec.~\ref{sec:DiscussionConclusions}.

\section{Materials and methods}\label{sec:materials_methods}
\subsection{The Molecular Dynamics simulation setup}

In the current study we used the Large-scale Atomic/Molecular Massively Parallel Simulator (LAMMPS) open-source software
package \cite{SandiaNationalLaboratories2015LAMMPSSandia,Bringuier2014MolecularExamples} to simulate the Cu self-sputtering process with different initial conditions. The Cu-Cu interactions were described by using the Embedded-Atom Method (EAM) interatomic potentials mainly by Mishin et al. \cite{Mishin2001StructuralCalculations}, but also by Mendelev et al. \cite{Mendelev2008AnalysisCu} to avoid the bias of results due to the use of a single potential. Both potentials  were splined to the Ziegler, Biersack and Littmark (ZBL) repulsive potential \cite{Ziegler2010SRIM2010}
at $r_{0}=1.5$ Å to take into account the short distance interactions correctly. The spline was done smoothly, so that none of the significant equilibrium parameters, which potentials the must reproduce, was affected by the ZBL potential.

We also compared our MD results of sputtering yields to the widely used semi-empirical formula by Yamamura et. al. \cite{yamamura1983theoretical,yamamura1983angular}, as well as to sputtering yields obtained by the binary collision approximation code SRIM \cite{Ziegler2010SRIM2010}, where all the interactions are described by only the ZBL potential and no thermal effects are taken into account.

Since we aim to consider the temperature dependence of sputtering yields, a special attention was paid to the value of the melting point, given by the potential. This value in the Mishin Cu potential was obtained by using the liquid-solid equilibrium approach \cite{Mor94,Nor97g}, resulting in the value of 1310$\pm$10 K.

\subsubsection{Electronic stopping power in MD simulations}

An energetic ion entering a material, loses its energy in both nuclear and electronic interactions. While the former is explicitly described by BCA or MD algorithms, the latter is not part of BCA or MD algorithms because of great difference in time scales (orders of magnitude) of both processes. However, it is not possible to obtain a physically meaningful result, if one of the energy loss channels is absent in simulations. There are different approaches to take into account electronic stopping power, a function describing interactions of energetic ions and atoms with electronic subsystem. One of them is implemented in LAMMPS as the two-temperature model, i.e. an energy exchange between nuclear and electronic subsystems of the materials, as the Langevin thermostat \cite{Duffy2007IncludingSimulations}. In the SRIM code, the electronic stopping is introduced as the work of a friction force, subtracted from the energy of the energetic atom between the collisions. A similar simplified approach is realized in the PARCAS MD code \cite{parcasshort}, which proved to provide simulation results, which are well compared to existing experiment \cite{Gre13,Ili13,Hol16}.

For the current simulations, we implemented the electronic stopping power using the same approach as in PARCAS. This allowed us not only to speed up the calculation of atomic cascades by LAMMPS, but also it enable the direct comparison to the sputtering yields obtained by the SRIM code and the classical semi-empirical formula by Yamamura, both widely used in the ion beam community.

Specifically, we apply an additional electronic stopping power in the form of a friction force to a projectile (an energetic atom) moving within the matrix with the kinetic energy exceeding a certain threshold (e.g. 10 eV by default). In other words, a decelerating force proportional to the atom velocity is subtracted from the force acting on the atom via other interactions in the direction of the movement. Moreover, this was calculated only if the energetic projectile had no less than 5 neighbors (the number of neighboring atoms on an arbitrary face of a FCC structure) to ensure that such decelerating force is applied to the projectile only within the target. This excluded
the possibility of electronic stopping power calculations in dimers and other small clusters that might be sputtered during the simulations. These new forces were calculated with the help of a fitted function of electronic stopping power values obtained from the SRIM database of experimental data \cite{Ziegler2010SRIM2010}: 
\begin{equation}
f[i][n]=f[i][n]-\text{sgn}(v[i][n])\cdot10^{0.217148\cdot \log(E_{k}\text{[eV]})-1.1887}\label{eq:Elstop}
\end{equation}
where $f[i][n]$ is the force acting on the $i$th atom in the $n$th
direction, $\text{sgn}(v{[}i{]}{[}n{]})$ is a signum function of the velocity
of the $i$th atom in $n$th direction and the constants used are the
fit parameters.

The latest version of electronic stopping power add-on for LAMMPS used the SRIM lookup tables
directly for faster calculations. This means that any element-in-element
electronic stopping power interaction approximations which are supported by SRIM can be
used by the electronic stopping power LAMMPS add-on. Nevertheless, Eq. \ref{eq:Elstop}
was tested in many situations and was found to produce physically
correct approximations for the electronic stopping power of Cu in Cu in the following simulations.

\subsection{Preparation of the simulation cell}

\begin{figure}
\centering
\includegraphics[clip,scale=0.5]{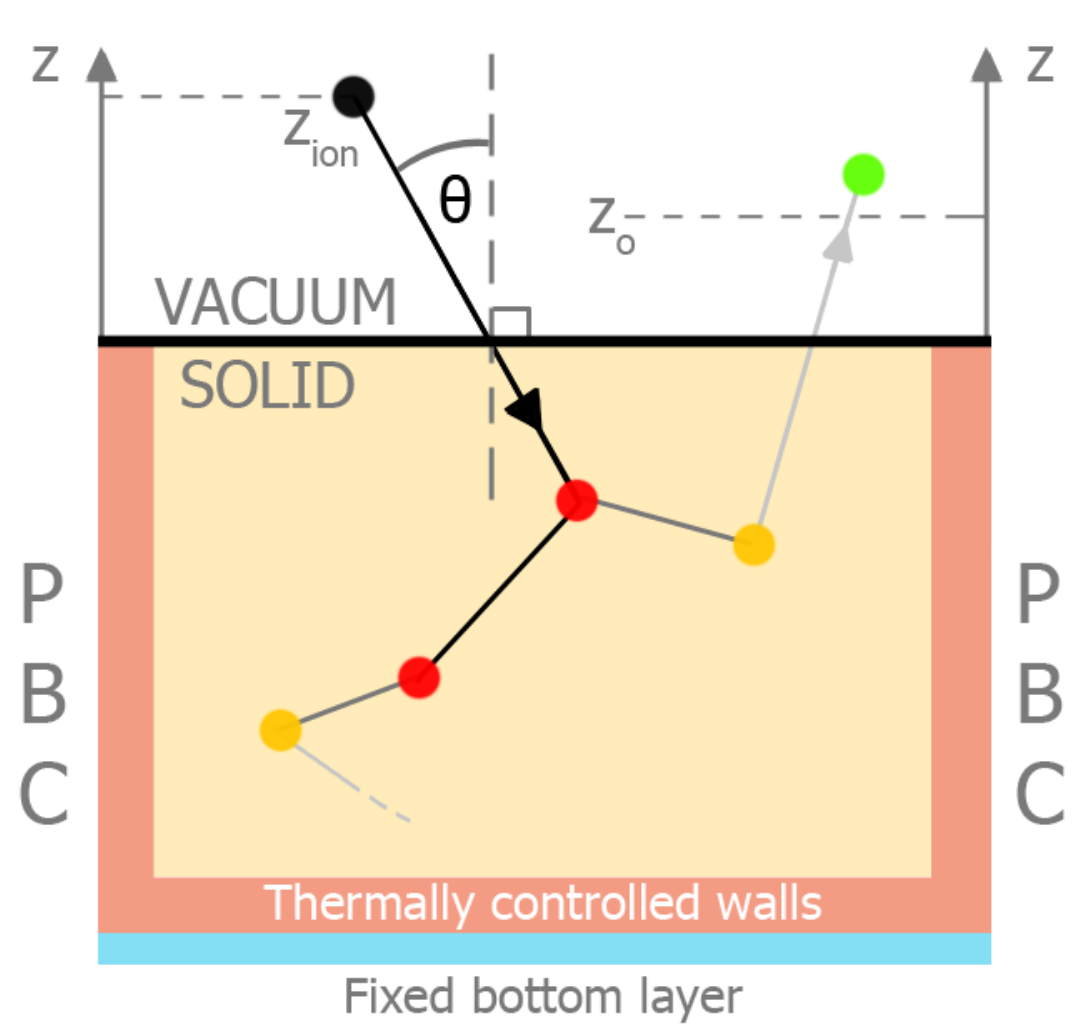} 
\caption{Schematic 2-dimensional illustration of the simulation setup (see the main text).}
\label{fig:MD-setup} 
\end{figure}

An illustration of the simulation box for our simulations of sputtering events is shown in Fig. \ref{fig:MD-setup}. The following procedure was used to prepare the structures. 

First the box of Cu atoms organized in the FCC lattice of a given size was relaxed in the NPT ensemble to allow for thermal expansion at the given temperature with periodic boundary condition in all three directions. 
The system was allowed to reach a stable configuration during 8000--20000 MD steps for each combination of target temperature, crystallographic orientation, and simulation box size. After this, the periodic boundary conditions were removed in the $z$-direction in order to make a free surface. The bottom atoms were fixed. This system was further relaxed in NVT.

Thereafter, thermally controlled boundary conditions were applied to a 1--2 unit cells thick layer at the sides and above the fixed layer at the bottom of the simulation box. The atoms in this layer were kept at the predefined temperature of the target to imitate the heat dissipation to the bulk, also preventing the overheating of the sample over periodic boundary conditions in the $x$- and $y$-directions.  To ensure that none of the important physical processes were affected by such boundary conditions, we tested several different sizes of the simulation box and monitored that the kinetic energy of the atoms reaching the thermally controlled boundaries is below $\sim$1 eV.

\begin{figure*}
\includegraphics[width=0.8\textwidth]{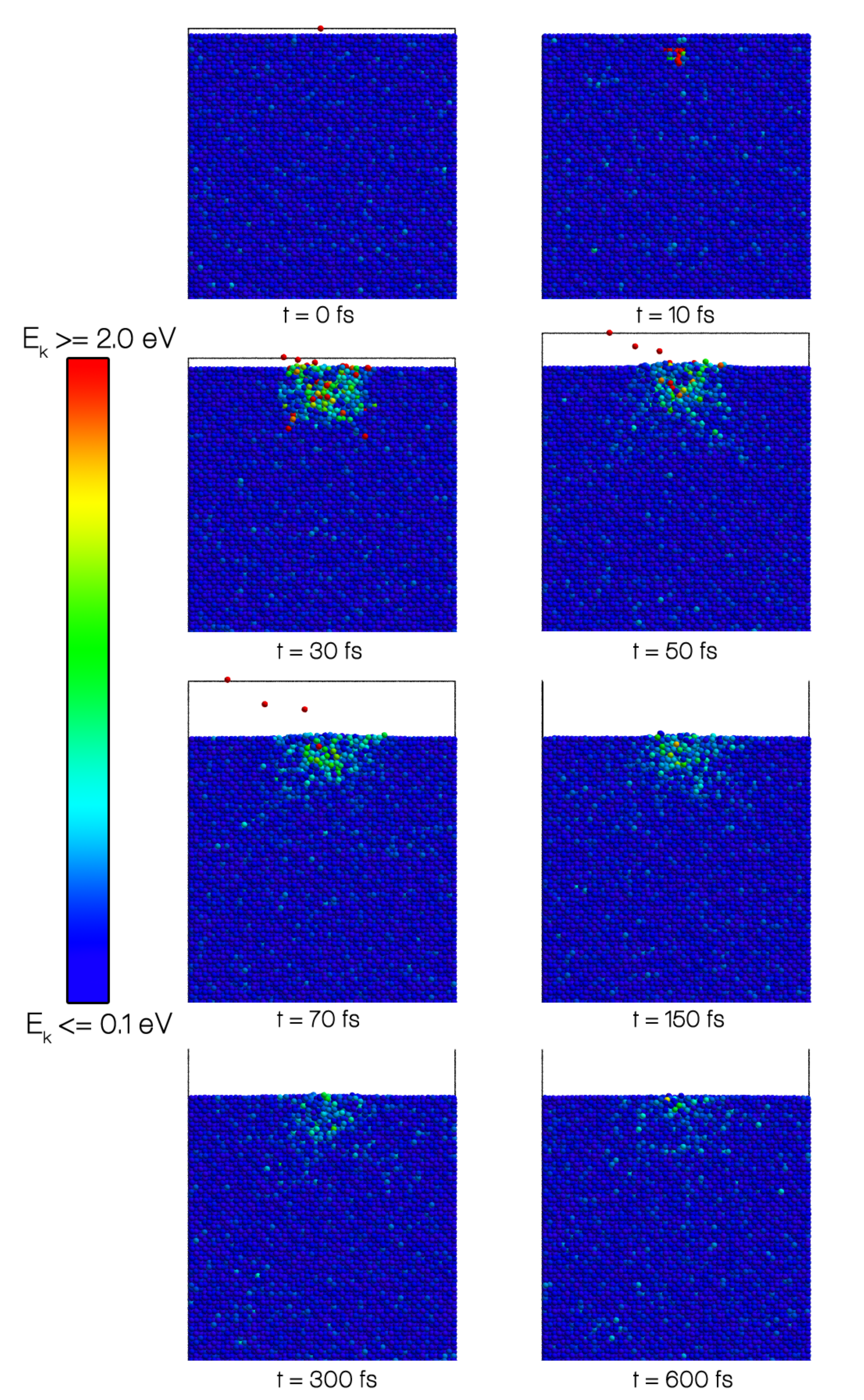}
\caption{Snapshots of cascade simulations, showing (using OVITO \cite{ovito}) the cross-sections at different time steps with the color scale giving the kinetic energies. The projectile kinetic energy is here $E_{k}=1$ keV and $\theta=3\degree$ on a \{100\} surface. The sputtering yield is 3.}
\label{fig:ADD-setup} 
\end{figure*}

The following sizes of the simulation boxes were used: $20\times20\times20$ unit cells for the projectile energies up to 1 keV, $34\times34\times40$ unit cells for 2 and 5 keV and $42\times42\times20$ unit cells for the simulations of angular dependence of sputtering yields. The latter box was expanded in the lateral directions to ensure that also cascades induced by ions incoming at glancing angles fit in the simulation cell. To illustrate the dissipation of energy from a 1 keV cascade, we show the principal stages of cascade development in Fig. \ref{fig:ADD-setup} with the color scale showing the kinetic energy of atoms.

In this study, we considered three different temperatures, 300 K (room temperature, operating
temperature of CLIC \cite{clic2016updated}), 500 K, 1200 K (areas around cathode spots on Cu surface heated up by field emission or vacuum arcing).

Finally, since MD simulations of ion bombardment deal with fast moving particles of high kinetic energies, an adaptive time step was introduced to avoid the problem of non-conserved total energy because of a too long distance the atom may move during a single time step. Each particle was allowed to move not further than 0.1 Å (magnitude of
atomic vibrations \cite{Cartz2002ThermalVibrations}) during one time step,
so the duration of the time step $\Delta t$ was reduced accordingly
within a range of $0.01 \leq \Delta t \leq 2$ fs. We checked that the energy of the system was conserved.

Each simulation of an ion impact was continued for approximately 2000--4000 time steps with 60--200 statistical runs depending on the kinetic energy
of the projectile.

\subsection{Initial inputs for simulations of sputtering events}

Initially the incoming Cu ion was positioned at 5--15 Å above
the topmost target layer near the center of the cell. The initial lateral position of the projectile was randomized within the half of a unit cell in both $x$- and $y$-directions to mimic the random impact of an ion on a crystal surface. The initial energy $E_{k}$ was chosen according to the CERN Fixed Gap System (FGS) experiment in the range from 100 eV to 5 keV \cite{Shipman2014ExperimentalCLIC}. Although the initial simulations were carried out for the full energy range,
eventually, the focus was shifted towards lower energies of 100--500 eV, since most of the ions, which are already in the low-temperature plasma regime (vacuum arcing) may reach the surface of the cathode exactly with this energy range \cite{Tim11b}. We also note that
the ions escaping plasma are accelerated mostly in the normal direction to the
surface as in CERN's FGS experiments \cite{Shipman2014ExperimentalCLIC}, however, we considered different incident angles to imitate a rough surface due to possible technological reasons or prolonged erosion during the plasma discharge.
The initial direction of the projectile was defined by an azimuthal (randomly selected between 0 and 2$\pi$) and a polar (incident) angles, $\phi$ and $\theta$, respectively.

We chose the angle of incidence, $\theta$, to have either a fixed value or a range of values to take into account possible spread of angles of incidence for each given MD simulation. The choices were motivated by the following:
\begin{itemize}
\item $0\degree \leq \theta \leq 3\degree$ --- to imitate normal incidence sputtering experiments (with 3 degree of uncertainty). Since these simulations were found to produce similar results,  $\theta=0$ was used whenever a normal incidence was studied;

\item $\theta$ = 7\textdegree{} or $\theta$ = 20\textdegree{} --- to reduce the effect of channeling, i.e. the phenomenon of abnormally deep penetration of incident ions due to a reduced stopping power in certain crystallographic directions \cite{Gemmell1974ChannellingCrystals,Yu1985AlignmentDeposition,Nor16};

\item $17\degree  \le \theta \le 23\degree$ or $27\degree \le \theta \le 33\degree$ --- to emulate rough features of $R \sim$1 nm in size;

\item $0\degree \leq \theta \leq 20\degree$ --- to emulate large scale roughness of surfaces ($R \sim 10$ nm);

\item From $\theta$ = 20\textdegree{} to $\theta$ = 85\textdegree{} --- to investigate the angular dependence of sputtering yield. 
\end{itemize}

\begin{figure*}
 \centering
 \subfigure[]{
  \includegraphics[width=0.3\textwidth]{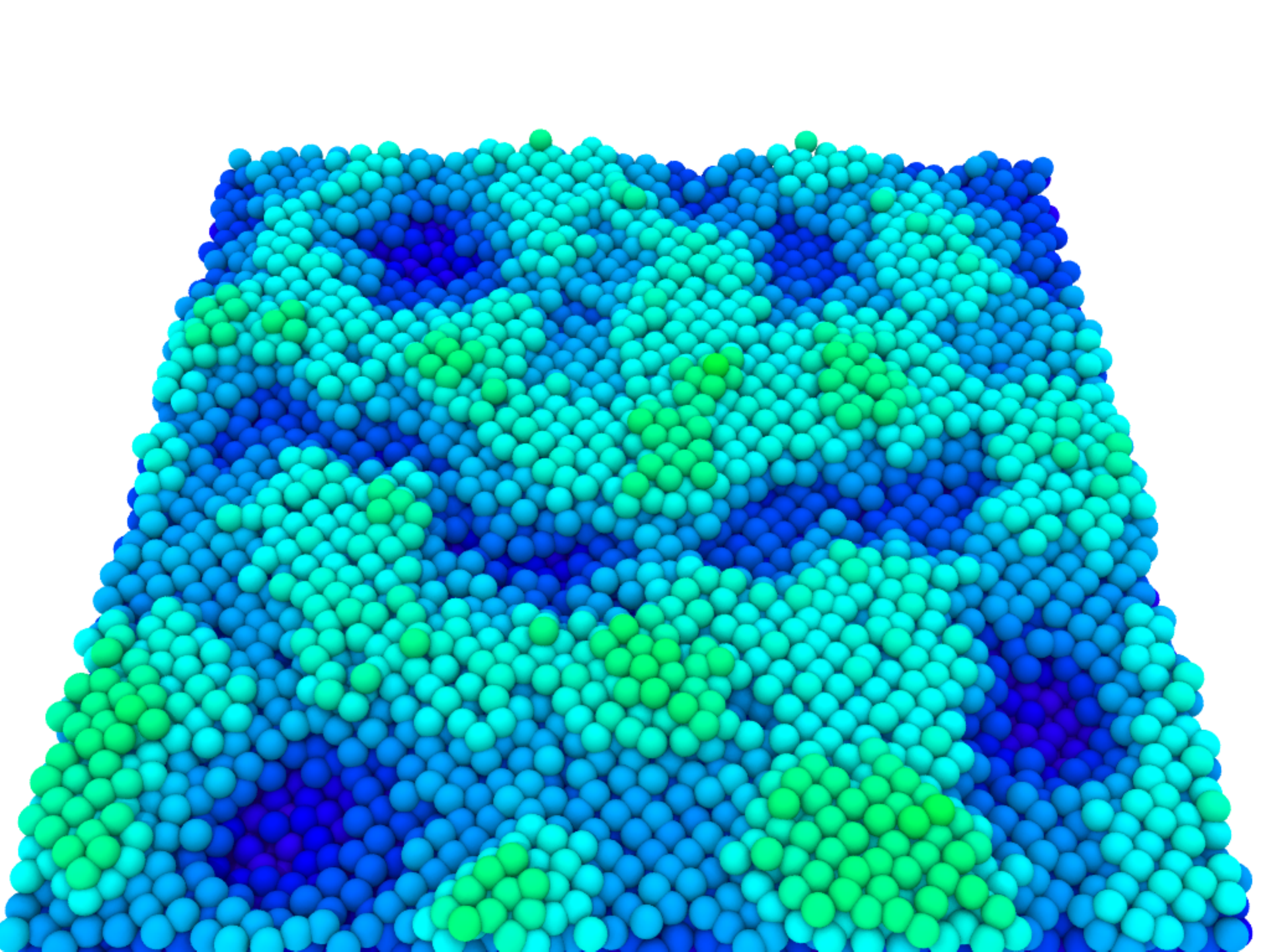}
 }
 \subfigure[]{
  \includegraphics[width=0.3\textwidth]{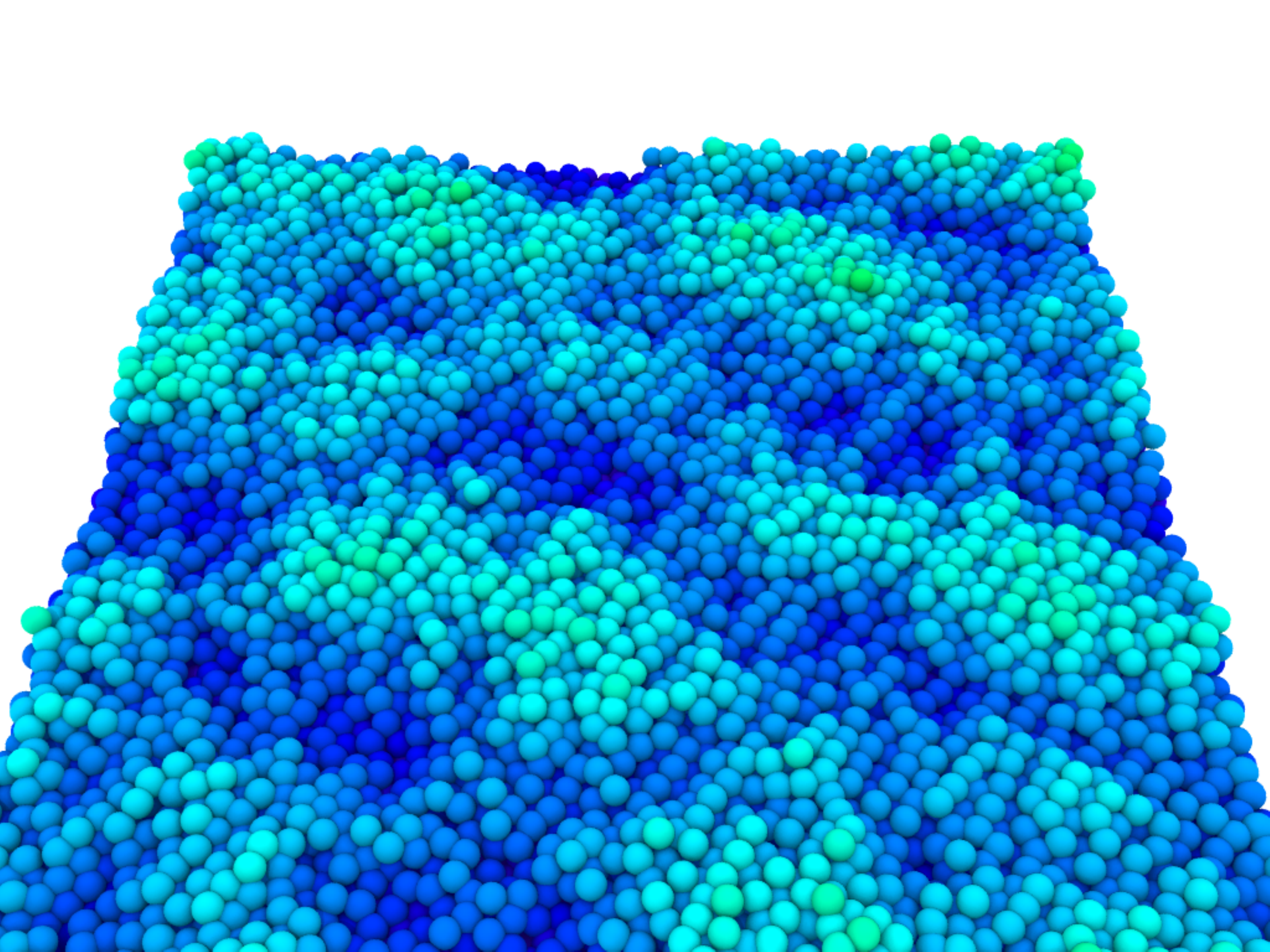}
 }
 \subfigure[]{
  \includegraphics[width=0.3\textwidth]{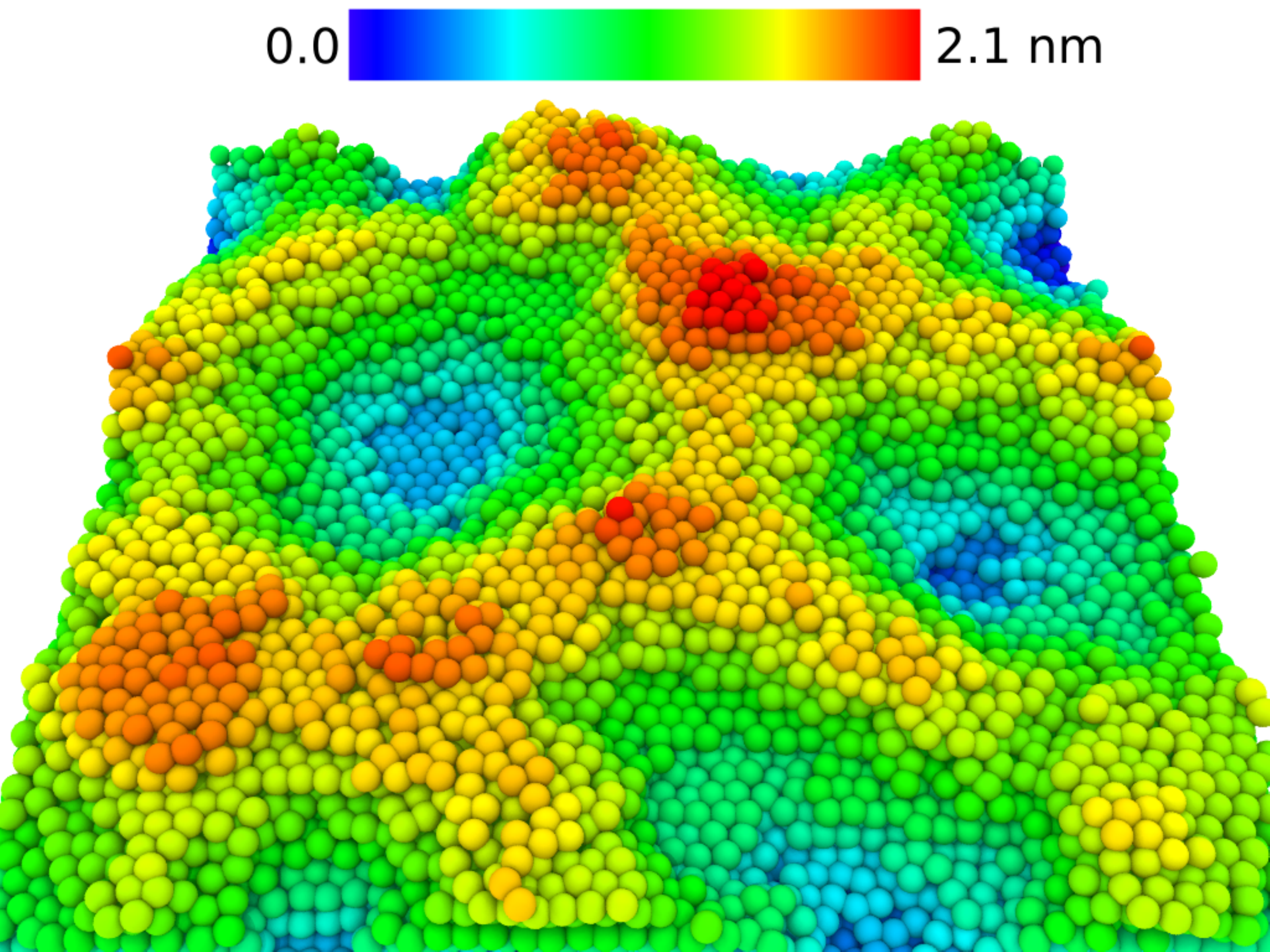}
 }
 \caption{Rough surfaces with the orientation (a) \{100\}, (b) \{110\}, and (c) \{111\}, constructed by randomly removing surface atoms in KMC and thereafter relaxed in MD at the target temperature of either 300 K or 1200 K (shown). The color scale shows the roughness in terms of height in nm, where the lowest surface point is set as 0.}
 \label{fig:KMC_surface}
\end{figure*}

We also simulated directly the surface roughness effects. Two different cases were considered in these simulations.
Surfaces with smaller roughness ($R\approx 1$ nm) were generated with the Kinetic Monte Carlo (KMC) simulation
code {\it Kimocs} \cite{jansson2016long}. The roughness was achieved by randomly removing surface atoms with an average rate of $10^{27}$ s$^{-1}$m$^{-2}$ at 300 K. The atoms were also allowed to diffuse, using a parameterization for surfaces, Cu Set 1, described in Refs. \cite{jansson2016long,baibuz2017migration}. Three different surfaces, \{100\}, \{110\} and \{111\} were used. The simulations were stopped after $10^7$ KMC steps or $\sim$$10^{-7}$ s. In order to be certain of the physical authenticity of these simulation cells, they were relaxed additionally in MD simulations. The snapshots of the final shapes of rough surfaces  generated by KMC method for all three basic orientations are shown in Fig. \ref{fig:KMC_surface}. We note here also that the irradiation of this surfaces was done slightly differently from a regular smooth surface. To provide the even ion coverage of the rough surface, we applied the algorithm of the random shift of the cell \cite{Nord01} --- the $(x,y)$ coordinates of all atoms in the cell were shifted by a random vector $(u_1L_x, u_2L_y)$ before every cascade simulation, after which the atoms that were moved outside the cell boundaries were shifted back in over the periodic boundaries. Here, $u_1$ and $u_2$ are the random numbers in the interval $]-0.5;0.5[$. This way the ion enters the surface always at the center to avoid the overlap with the boundaries, however, every time it "sees" a new surface with a different morphology. This trick allows to avoid the bias of the results from the entrance at the same position and possible artifacts from the overlap with periodic boundaries in lateral directions.    

Large scale surface roughness ($R \ge 10$ nm) was emulated by using the previously mentioned angular ranges of bombarding particles.

\subsection{Calculation of sputtering yields and ion ranges}

In our simulations, we focused on two major quantities characterizing the ion bombardment event: sputtering yield and penetration depth or ion range. All atoms found at the end of each simulation above $z_{0}=5$ Å above the surface level (the highest initial $z$-coordinate of an atom in the topmost surface layer) were counted as sputtered and contributed to the sputtering yield. The angular dependence of this quantity obtained at room temperature was compared to the semi-empirical formula by Yamamura \cite{yamamura1983angular} to test the applicability of such formulas for the general case of various material conditions.

The ion range was calculated as the difference between the final $z$-coordinate of the projectile and the average $z$-coordinate of the topmost atom layer.

\section{Results}\label{sec:Results}

\subsection{Cu self-sputtering yields at different incident angles at room temperature}\label{sec:angular}

\subsubsection{Sputtering yields at normal incidence}

\begin{figure}
 \includegraphics[width=\columnwidth]{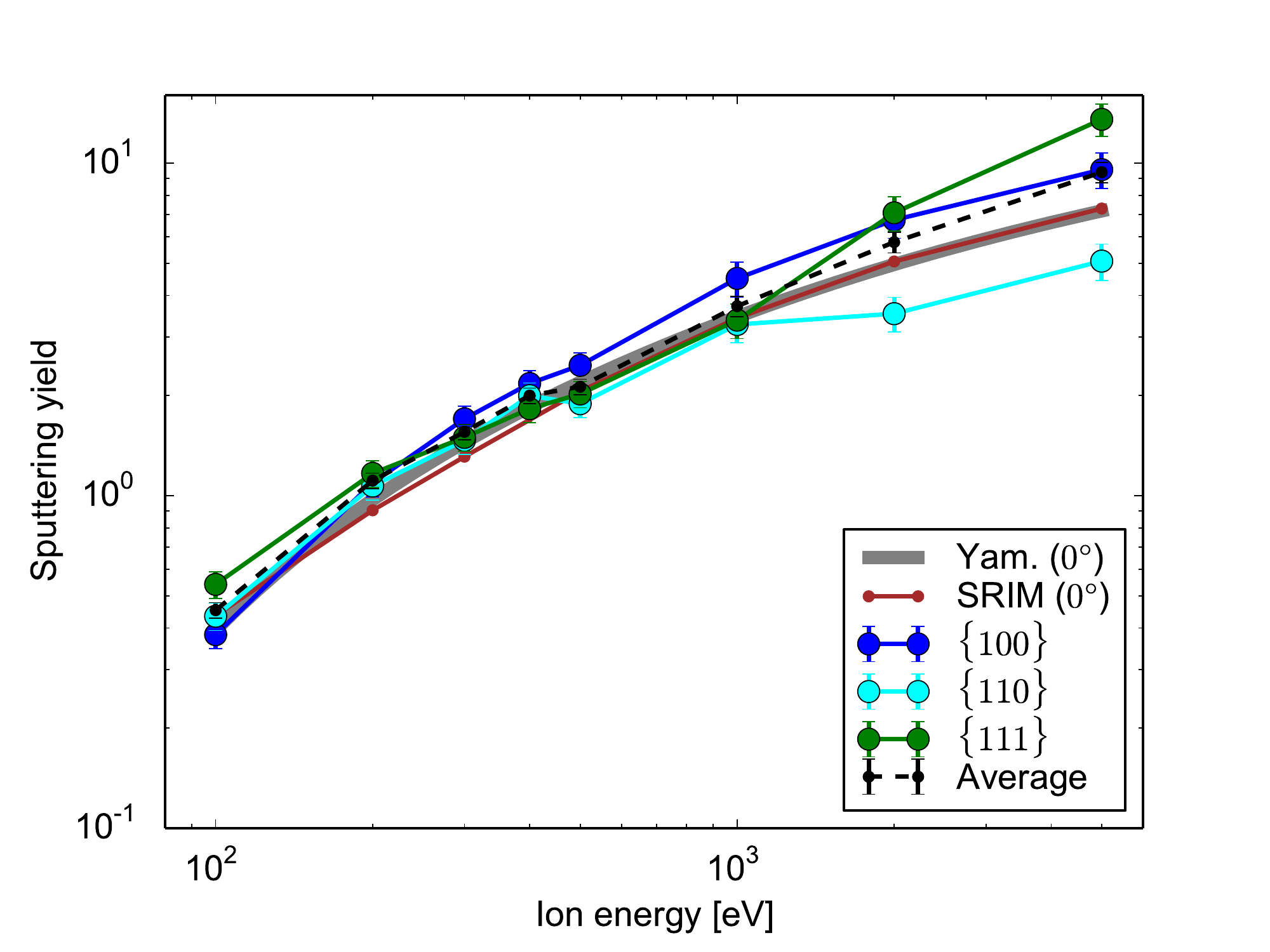}
 \caption{Sputtering yields obtained at 300 K for normal incidence from three different surface orientations as well as the polycrystal surface (weighted average sputtering yield) are plotted versus incoming ion energies. The results
from SRIM calculations as well as the semi-empirical model by Yamamura et al. \cite{yamamura1983theoretical,matsunami1984energy} (Yam.) is shown for comparison.} 
\label{fig:orient1} 
\end{figure}

The sputtering yields obtained in our simulations for normal incidence for each crystallographic face (including the polycrystal surface) are presented in Fig.~\ref{fig:orient1}. For comparison, included are also the yield obtained by using the formula by Yamamura \cite{yamamura1983theoretical} with fitting parameters from \cite{matsunami1984energy} as well as the yield calculated by SRIM \cite{Ziegler2010SRIM2010}.
It can be seen that for normal incidence, the MD results are consistent with the Yamamura's and SRIM sputtering yields.
We note that the sputtering yields for the ion energy $>$ 1 keV for the \{110\} surface is clearly lower than that obtained for other surfaces or by Yamamura's formula or SRIM calculations. Most likely this result is due to the channeling effect, which is strong in the $\langle110\rangle$ directions parallel to the close-packed \{111\} crystallographic planes in the FCC lattice.

\subsubsection{Sputtering yields under tilted incidence}

\begin{figure}
\includegraphics[width=\columnwidth]{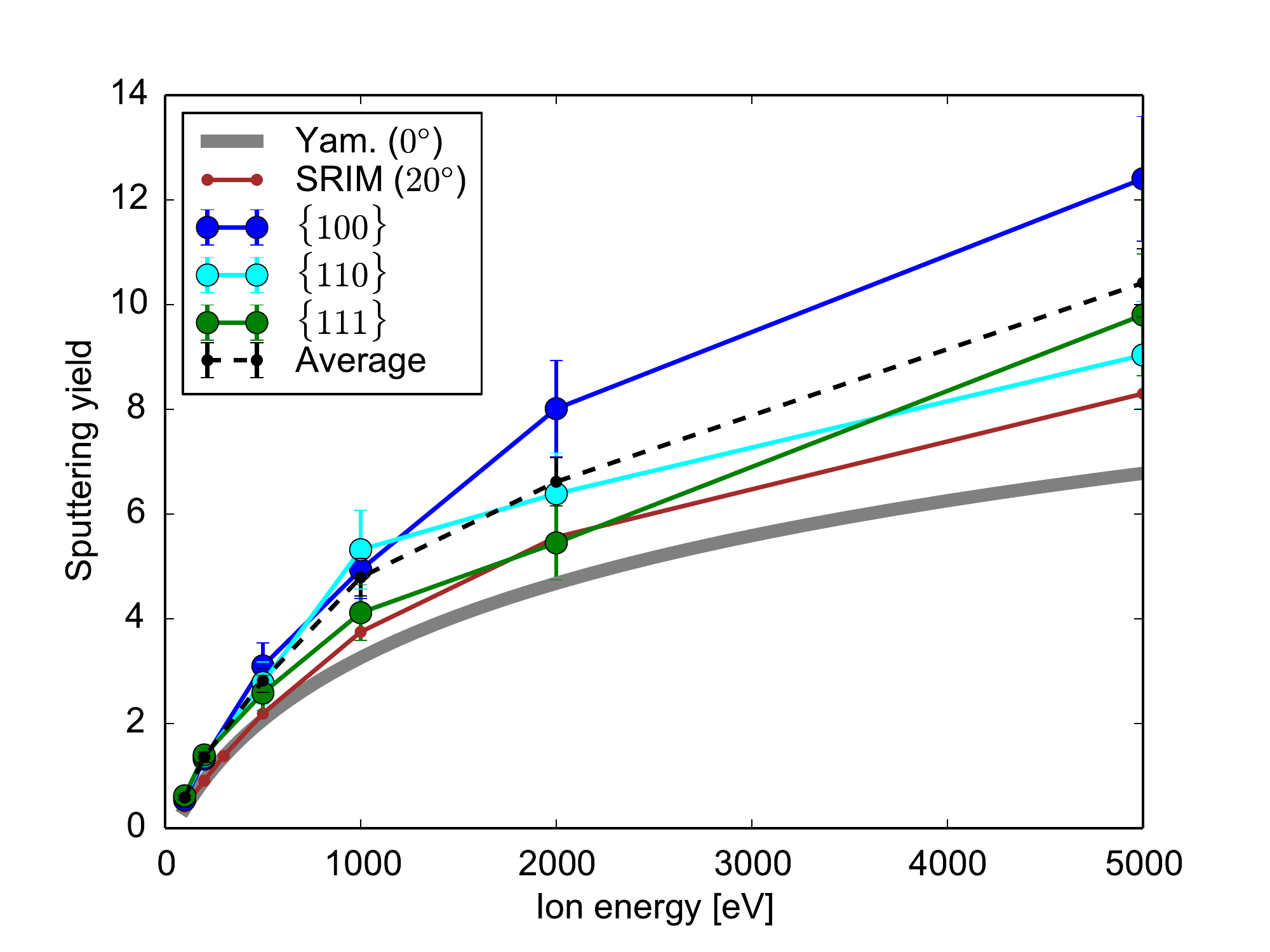}
\caption{Sputtering yield at 300 K for $\theta=20\degree$ with different crystallographic
orientations and their average. The Yamamura sputtering yield fit ($\theta=0\degree$) \cite{yamamura1983theoretical,matsunami1984energy} and SRIM sputtering yield data ($\theta=20\degree$) are included for comparison. 
}
\label{fig:orient2}
\end{figure}

In order to reduce the channeling effect \cite{Nor16,Gre13}, we have selected two fixed values of the incident angles: $\theta=7\degree$
and $\theta=20\degree$ \cite{Gemmell1974ChannellingCrystals,Yu1985AlignmentDeposition}. As expected, the simulations at 300 K and $\theta=7\degree$ resulted in similar behavior of $Y(E_k)$ as in Fig.~\ref{fig:orient1}
(although the sputtering yield for the \{110\} surface slightly increased due to reduced channeling), so we do not show them here. 
The results obtained with $\theta=20\degree$ and at 300 K are shown in Fig.~\ref{fig:orient2}. Here we also show $Y(E_k)$ from SRIM simulations with the same tilted incidence and the Yamamura's formula, which does not have an angular dependence and thus the result is given for normal incidence. Under this condition, we see that the MD simulations resulted in higher sputtering yields, compared to both analytical model and BCA simulations. Moreover, the highest $Y(E_k)$ we obtained for the \{100\} surface.

\subsubsection{Sputtering yields by ion impacts spread within a solid angle}

\begin{figure}
\includegraphics[width=\columnwidth]{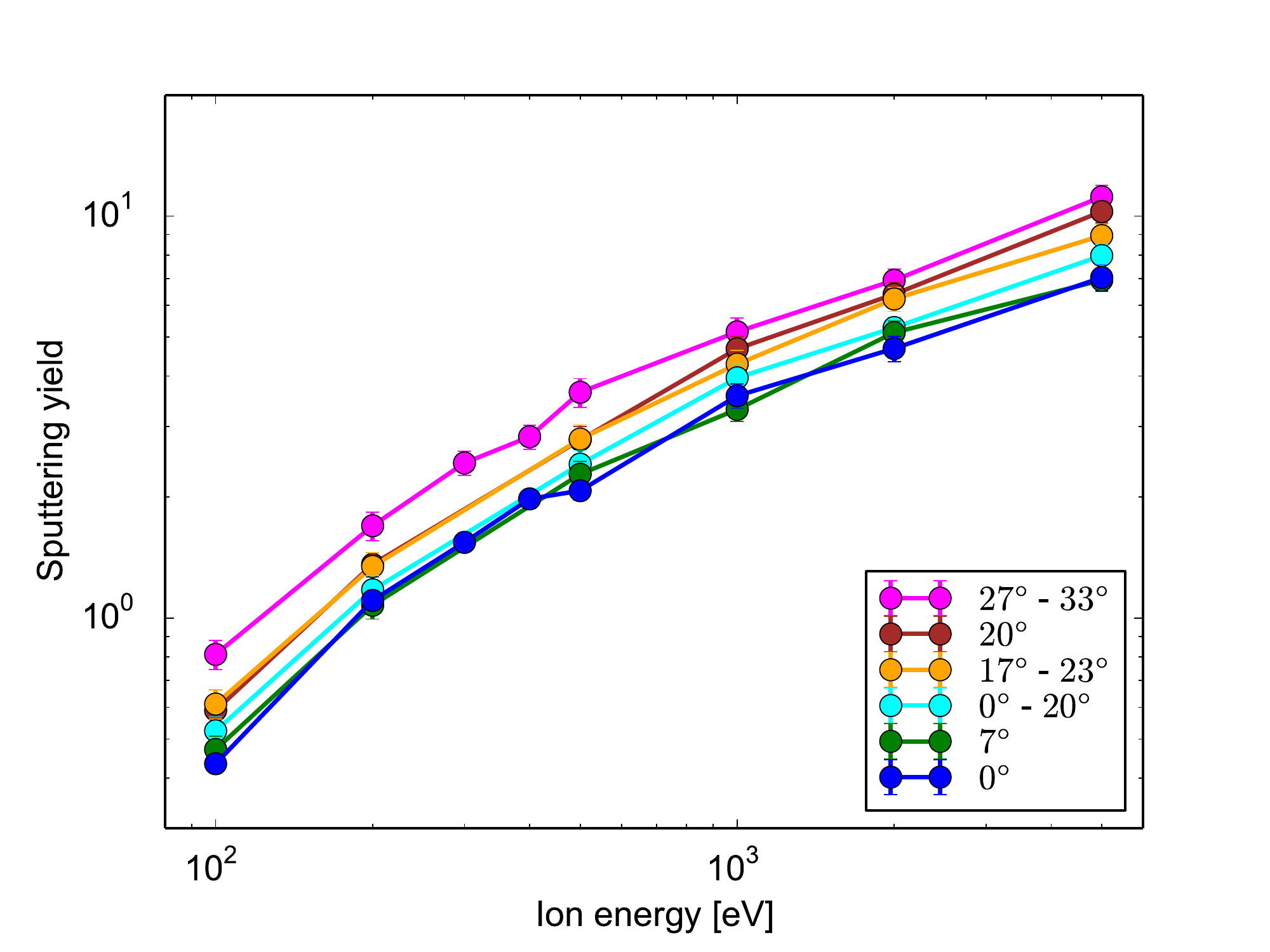}
\caption{Averaged sputtering yields for surfaces held at 300 K for different angles of incidence.}
\label{fig:angle2} 
\end{figure}

For better understanding of the angular dependence of sputtering yields in realistic condition of a vacuum arc, we performed a set of simulations where the $\theta$ angle was not fixed, but was allowed to vary gradually within different intervals of angles (a solid angle) for all energies at $T=300$ K. These results are presented in Fig.~\ref{fig:angle2}. All simulations of this kind were performed for the \{100\} surface. If we compare the results obtained for the fixed angle and the solid angle around this fixed value, we do not see a big difference, the curves practically lie on top of each other. However, the expected overall tendency of increasing of the sputtering yield with increase of the angle of incidence is clear for both fixed (exact)  and the solid (with a spread $\pm 3\degree$) angles. 

\subsection{Analysis of channeling}\label{sec:channeling}

As we saw in section \ref{sec:angular}, channeling may play a significant role for the self-sputtering yield of Cu, depending on crystallographic orientation of the surface. Since the grains of a polycrystalline surface may be oriented in a random direction, we analyze here in which directions the channeling effect is sufficiently strong in Cu lattice and what is the energy range when the channeling effect becomes relevant. We used the systematic approach introduced in
Ref. \cite{Nor16} for analyzing channeling over all crystal
directions. In the current case, we simulated the mean ion range of
100 eV -- 5 keV Cu ions on a
(001) surface with the MDRANGE method for polar angles $\theta$ from 0$\degree$ to 89$\degree$, for azimuthal angles $\phi$ from $0\degree$ and $90\degree$.  For the energies of 500 eV and 5 keV we also simulated the entire channeling map over all directions.

\begin{figure*}
\centering 
\subfigure[]{
\includegraphics[width=\columnwidth]{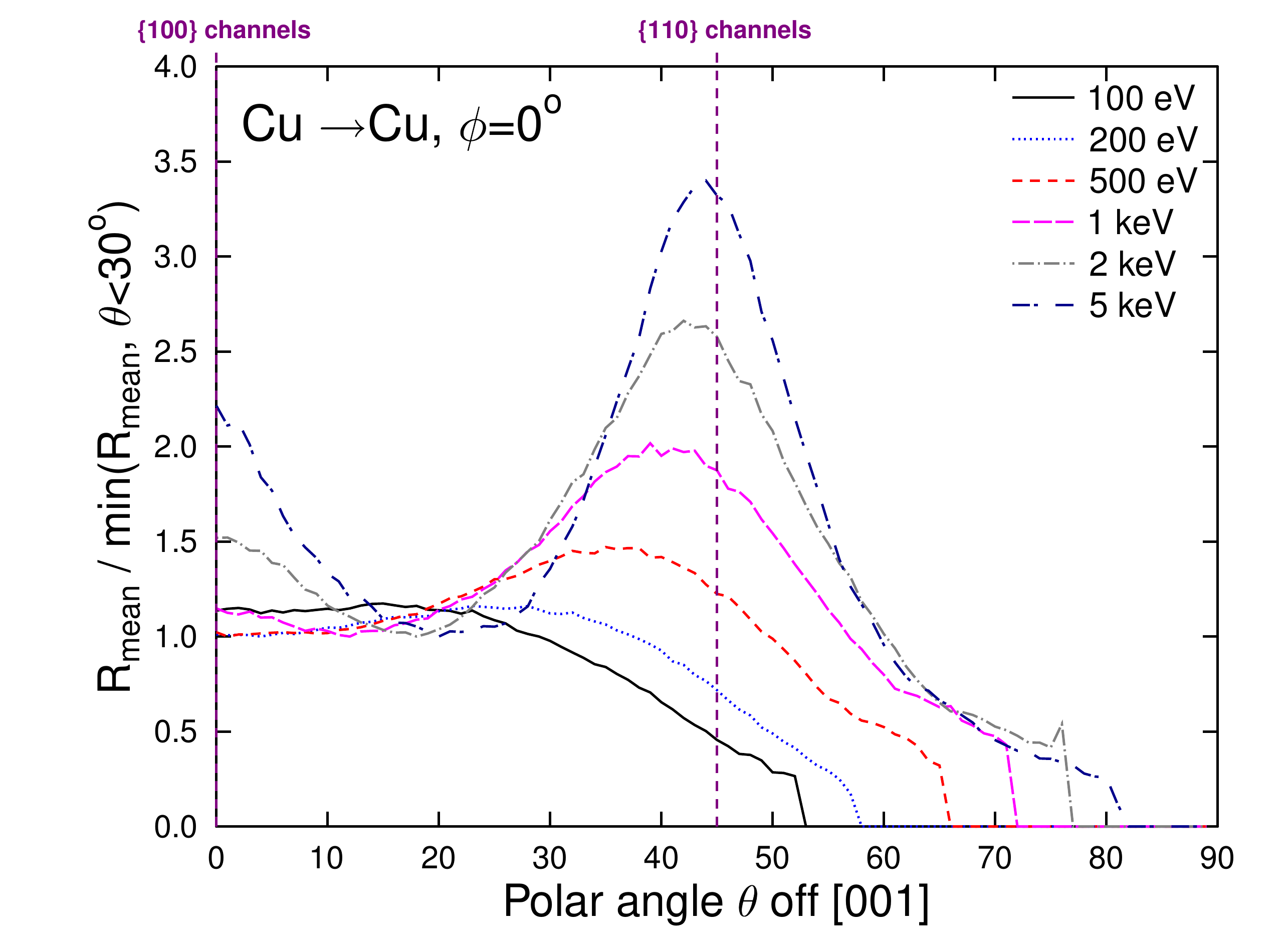}
}
\subfigure[]{
\includegraphics[width=\columnwidth]{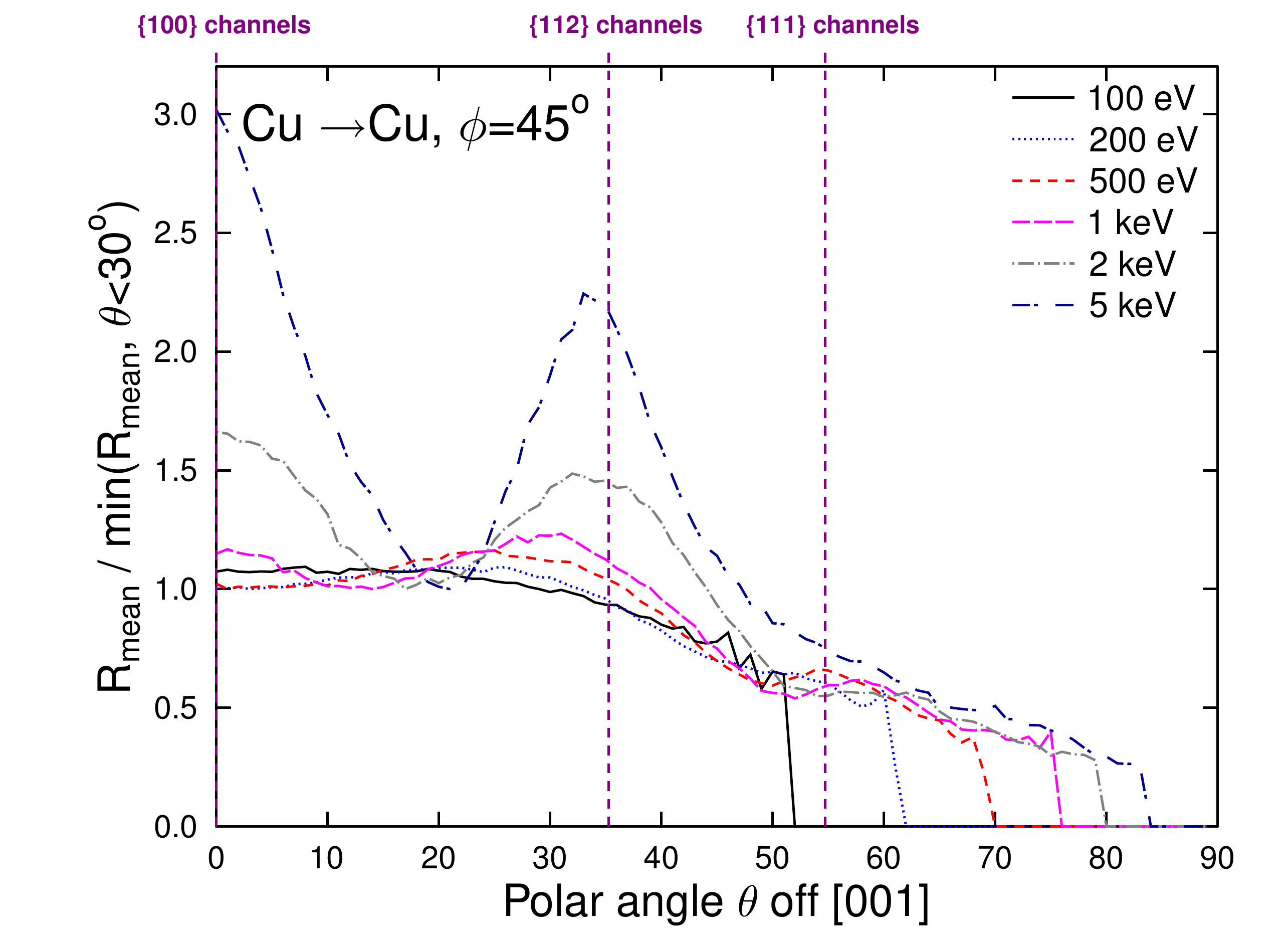}
}
\caption{
Mean ion ranges for Cu ions in Cu as a function of the 
polar (tilt) angles off $[001]$ for azimuthal angles 
of a) $\phi=0^\circ$ and b) $\phi=45^\circ$.
Note that due to planar channeling effects, the normalization
factor $\min(R_{\rm mean}, \theta<30^o)$ is somewhat different
for $\phi=0^\circ$ and $\phi=45^\circ$.
}
\label{fig:channeling}
\end{figure*}

The results on the mean ranges for different energies are illustrated 
in Fig. \ref{fig:channeling}. For clarity on the relative importance of channeling, the mean ion ranges $R_{\rm mean}$ are plotted normalized with the minimum value
of the mean range below $\theta=30\degree$. 

\begin{figure*} 
\centering
\subfigure[]{
\includegraphics[width=\columnwidth]{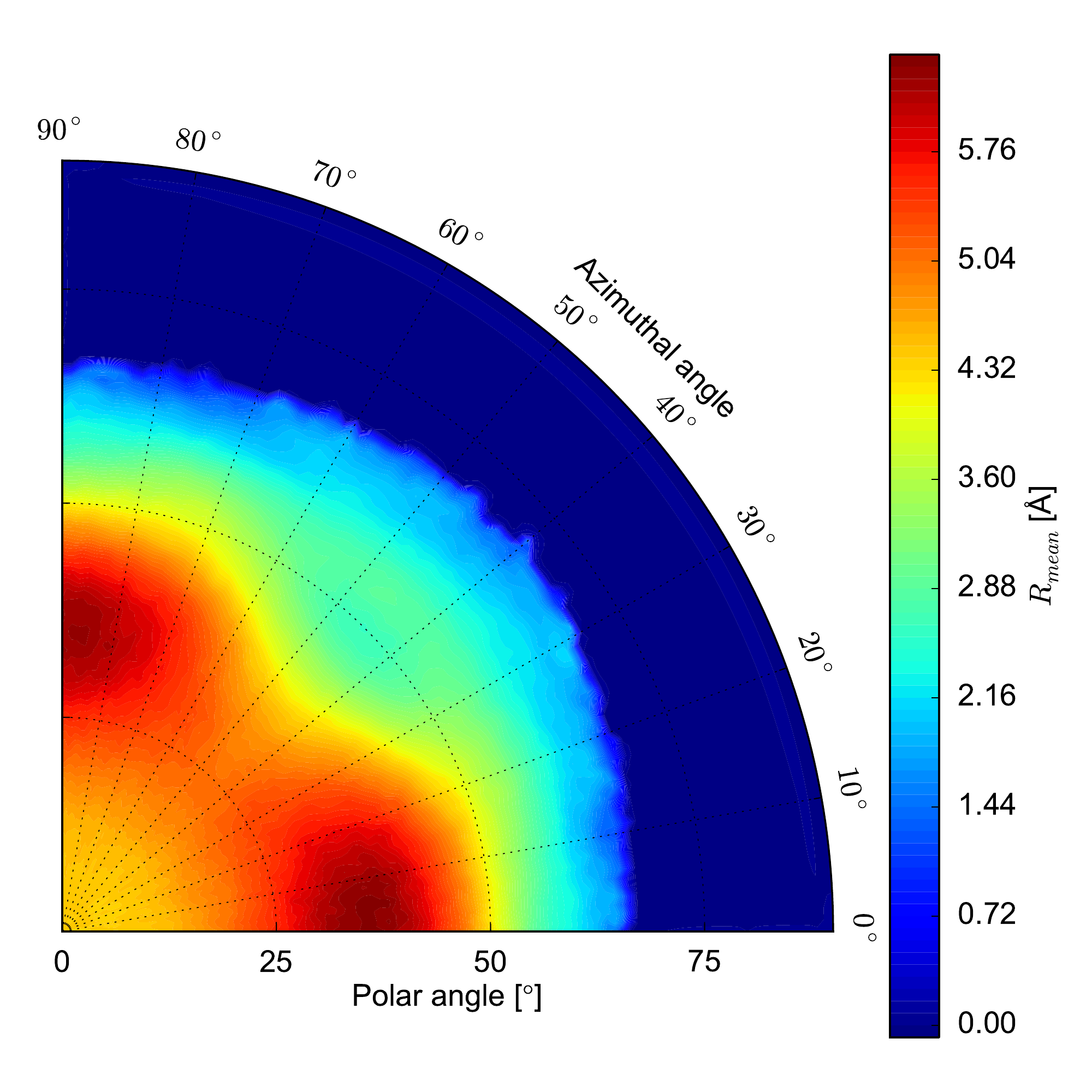}
}
\subfigure[]{
\includegraphics[width=\columnwidth]{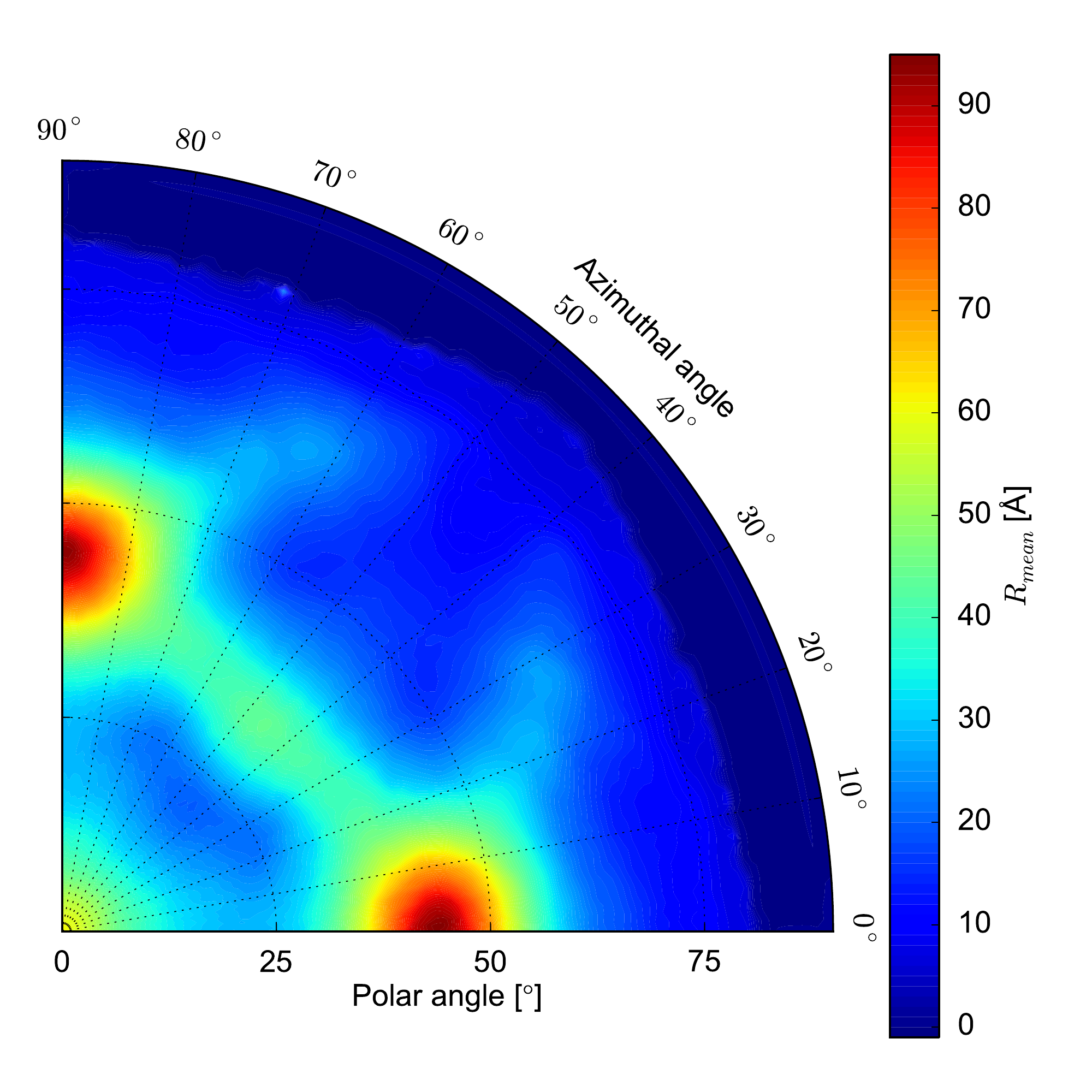} 
}
\caption{
Channeling maps of mean ion ranges for (a) 500 eV 
and (b) 5 keV Cu ions in Cu. In the dark blue areas, all
ions were reflected off the surface and hence no mean range 
could be defined.}
\label{fig:channelingmaps}
\end{figure*}

The results in the figure show that below 500 eV, there is no significant channeling in any crystal direction. At 500 eV, there is a minor enhancement of ion ranges near the $\{110\}$ directions, see Fig. \ref{fig:channelingmaps}a. The maximum range is slightly below the principal direction due to steering effects \cite{Nor16}.

  Channeling starts to be significant around $\gtrsim 2$ keV in the $\{100\}$ directions,
around $\gtrsim 1$ keV in the $\{110\}$ directions, and
is not significant at all in the $\{111\}$ directions in the current range
of energies studied. 
The  $\{110\}$ direction exhibits by far the strongest
channeling as expected from previous results in other FCC
metals systems \cite{Nor16}.

  These MDRANGE results on channeling explain the full MD (LAMMPS) results on sputtering, which showed that the sputtering yield is significantly affected by channeling only in the $\{110\}$
directions.

\subsection{Temperature dependence of the self-sputtering yields}\label{sec:temperature}

\begin{figure*}[ht!]
\centering
\subfigure[]{
\includegraphics[width=\columnwidth]{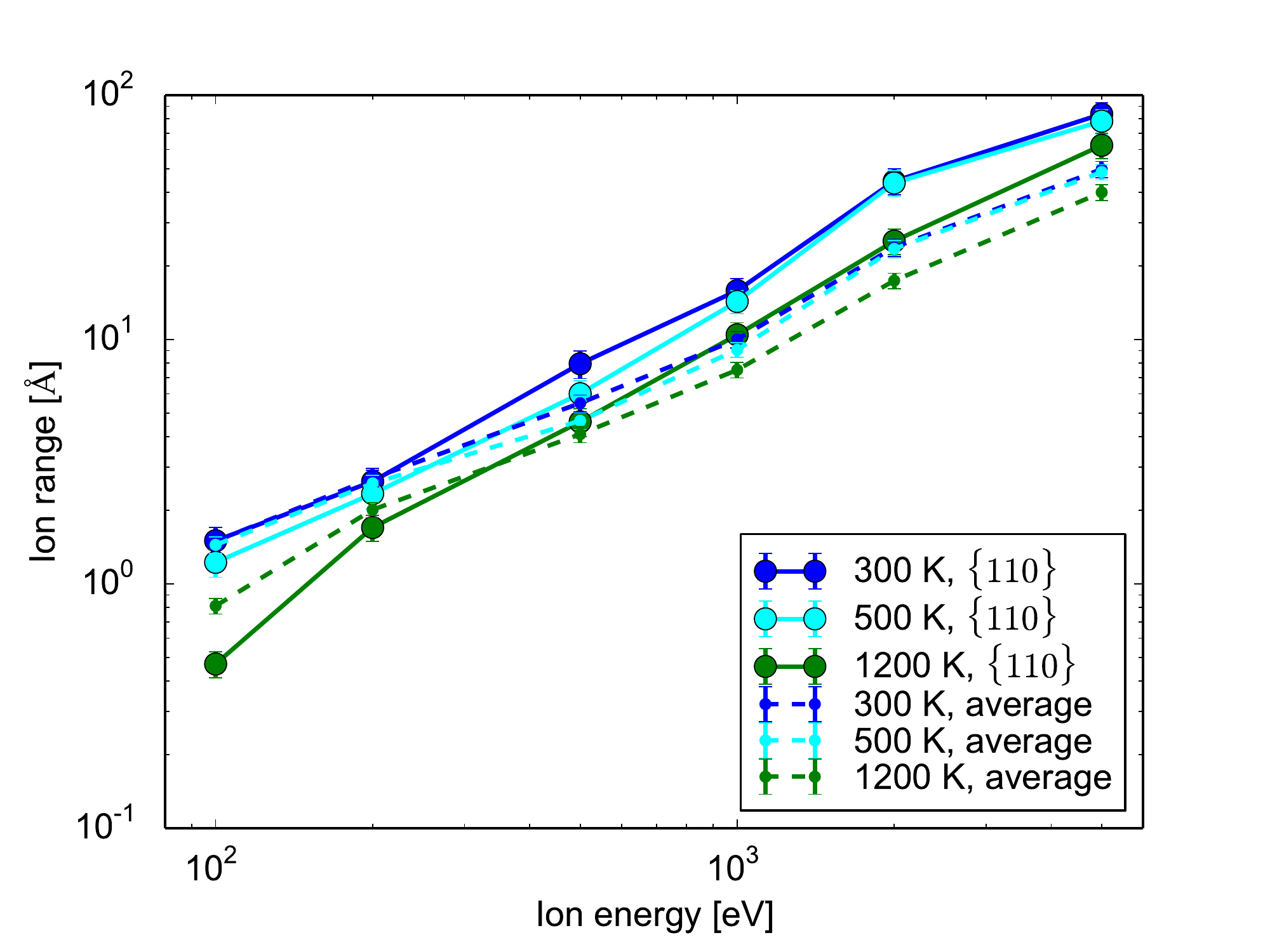}
}
\subfigure[]{
\includegraphics[width=\columnwidth]{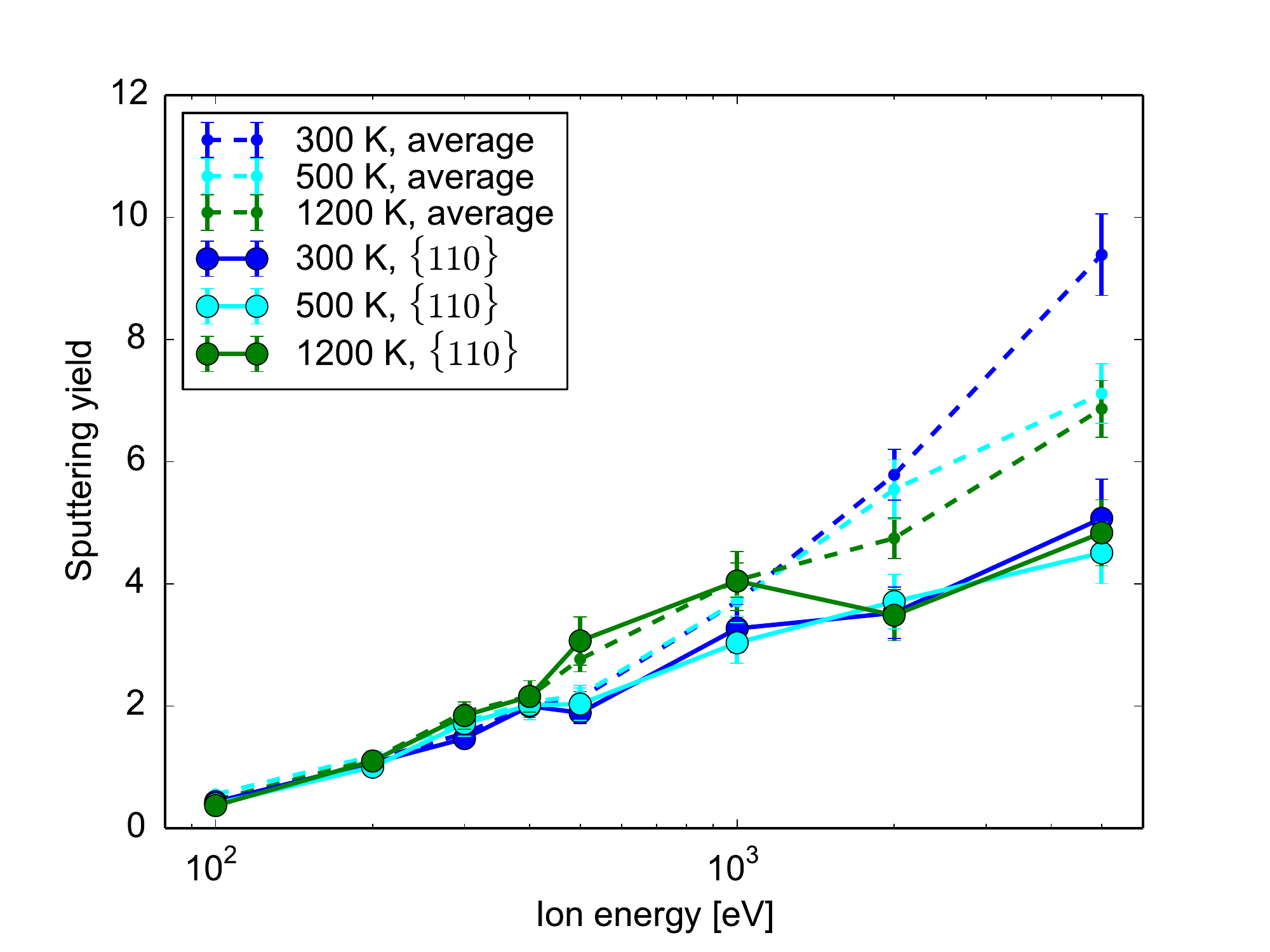}
}
\caption{Ion range (a) and sputtering yield (b) as functions of the incident energy obtained for the \{110\} surface at the normal incidence at different temperatures. The average values are over the three different surfaces.}
\label{fig:orient4}
\end{figure*}

We performed the series of simulations at 300 K, 500 K and 1200 K for the ion energies 100 eV -- 5 keV to analyze the temperature dependence of self-sputtering yields of Cu. Since increase of temperature enhances atomic vibrations in crystal structures, we expect the largest effect of the temperature when the ion enters the lattice along the crystallographic directions with the strongest channeling effect. We analyzed the dependence of the ion range (penetration depth) on the temperature calculated for the \{110\} surface shown in Fig.~\ref{fig:orient4}a. For comparison, we also show the ion ranges averaged over the three surfaces for different temperatures, approximating the ion range for a polycrystalline target. The \{110\} data has higher ion range for high ion energies, but lower than the average data at very high temperatures, and especially for the low energy ions, since the atomic vibrations in the lattice sites then become sufficiently significant to reduce the channeling in this direction. In Fig.~\ref{fig:orient4}b, the energy dependence $Y(E_k)$ is shown for the \{110\} surface as well as for the polycrystalline surface at different temperatures. 

We do not show here the results for \{100\} and \{111\} surfaces, since we did not observe significant difference of the results with increase of temperature -- all results for the ion range and the sputtering yields were similar to the reported dependencies in Figs.~\ref{fig:orient4}a and \ref{fig:orient1}, respectively.

\begin{figure}
\includegraphics[width=\columnwidth]{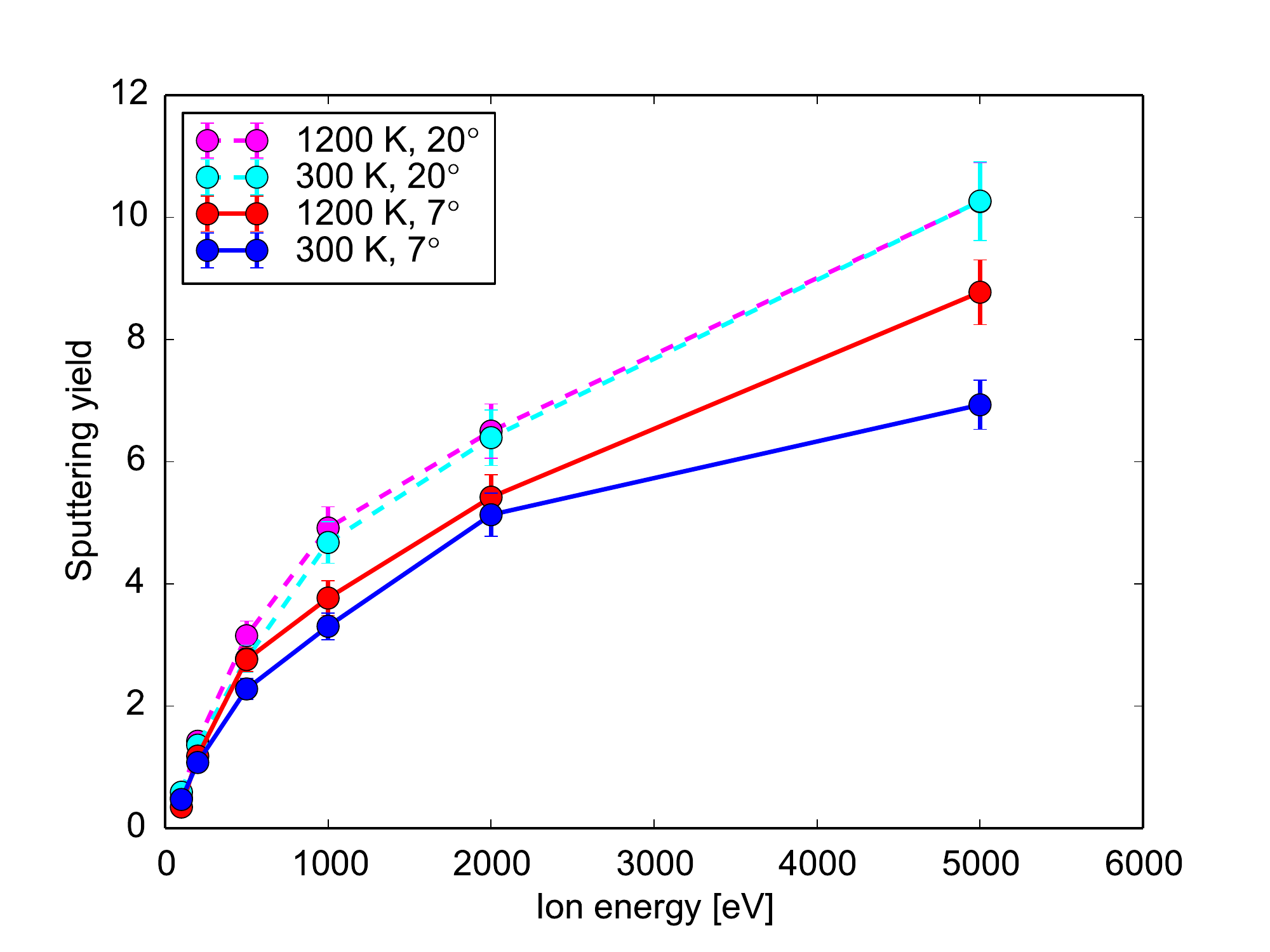}
\caption{Sputtering yield of Cu from the \{100\} surface irradiated at 300 K and 1200 K with $\theta=7\degree$ and $\theta=20\degree$ of incident angles.}
\label{fig:angle1} 
\end{figure}

The comparison of the sputtering yields for \{100\} surface for 300 K and 1200 K with incident angles of $\theta=7\degree$ and $\theta=20\degree$ are shown in Fig.~\ref{fig:angle1}. 
We see that the sputtering yield increased slightly with increase of temperature for $\theta=7\degree$, but for $\theta=20\degree$, the
difference is not distinguishable.
An impinging ion with a steep incident angle, like 7\textdegree, has a high probability for a deep penetration of the \{100\} surface due to the channeling effect, as seen in Fig. \ref{fig:channeling}, which means that less of the surface atoms will be affected by the cascade and sputtered. This is especially true at low temperatures like 300 K. At higher temperatures, like 1200 K, the ion range is lower, as is also seen in Fig. \ref{fig:orient4}. With the larger incident angle of 20\textdegree, the channeling effect and the ion range will be small independently of the temperature (Fig. \ref{fig:channeling}).

\begin{figure}
\includegraphics[width=\columnwidth]{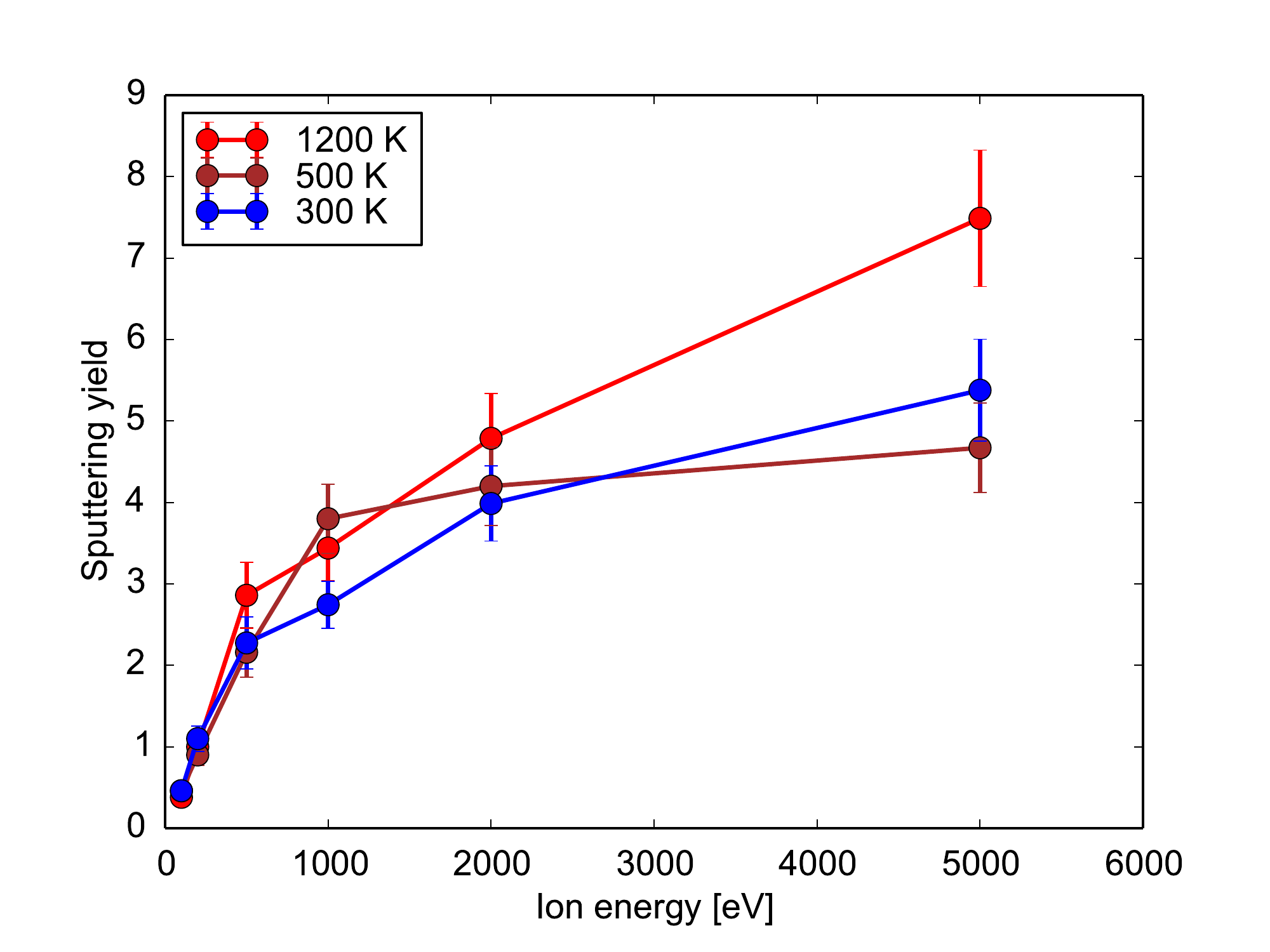}
\caption{Sputtering yield of Cu from the \{110\} face with $\theta=7\degree$ at different temperatures.}
\label{fig:orient3} 
\end{figure}

The most different sputtering yield with increased temperature we observed for the surface oriented in the \{110\} direction held at the temperature 1200 K with $\theta=7\degree$  and $>$ 1 keV energy of the incident ion (Fig.~\ref{fig:orient3}). This difference, however, did not exceed 50 \%. We observed the difference only for very high temperature, while the difference in $Y(E_k)$ between 300 and 500 K is rather within the statistical uncertainty related to the corrugated nature of \{110\} surfaces.

\begin{figure}
\includegraphics[width=\columnwidth]{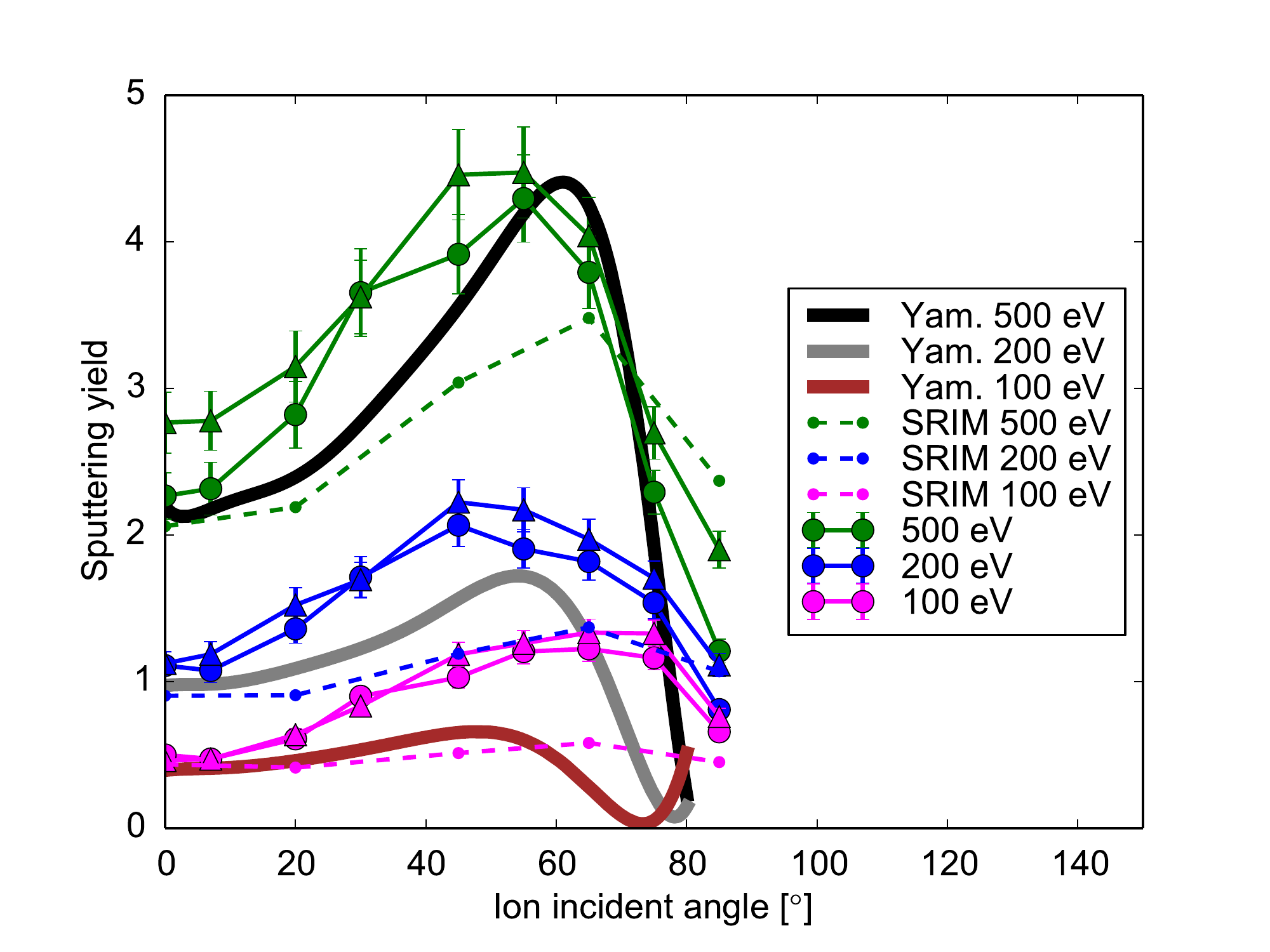}
\caption{Averaged sputtering yield versus $\theta$ at 300 K (bullets) or 1200 K (triangles) for different ion energies. SRIM calculations and angular dependent sputtering yield from Yamamura's formula \cite{yamamura1983angular} are added for comparison.}
\label{fig:angle4} 
\end{figure}

For the low ion energies, i.e. 100 eV, 200 eV and 500 eV, we also analyzed the full angular dependence of $Y(\theta)$ for two target temperatures 300 and 1200 K.
These results summarized in Fig.~\ref{fig:angle4} are in line with experimental observations for Cu \cite{Fluit1963Angular-dependentCrystals,Weh59,Yam83}.  In this graph, we see that some effect of temperature on sputtering yield can be seen only for the highest studied energy, 500 eV. However, the difference is rather within the statistical uncertainty of the obtained results.

A close inspection of these results reveals that at the incident energies from 200 eV and up, the maximum of the sputtering yield is shifted in MD results towards the smaller incident angles compared to both Yamamura and SRIM results. However, in the same comparison at the lowest energy of 100 eV, the SRIM results compare much better to the MD (the positions of the peaks of distributions), while the peak in the Yamamura results is shifted towards smaller incident angles. It is clear that while at higher energies the Yamamura's formula captures the physics of sputtering process fairly well, it fails to describe the angular dependence of sputtering yield at low energies, when many-body interactions become essential. Although these are not present in BCA algorithm of SRIM calculations either, the quantitatively underestimated sputtering yields still exhibit a angular dependence as our MD results.

\begin{figure}
\includegraphics[width=\columnwidth]{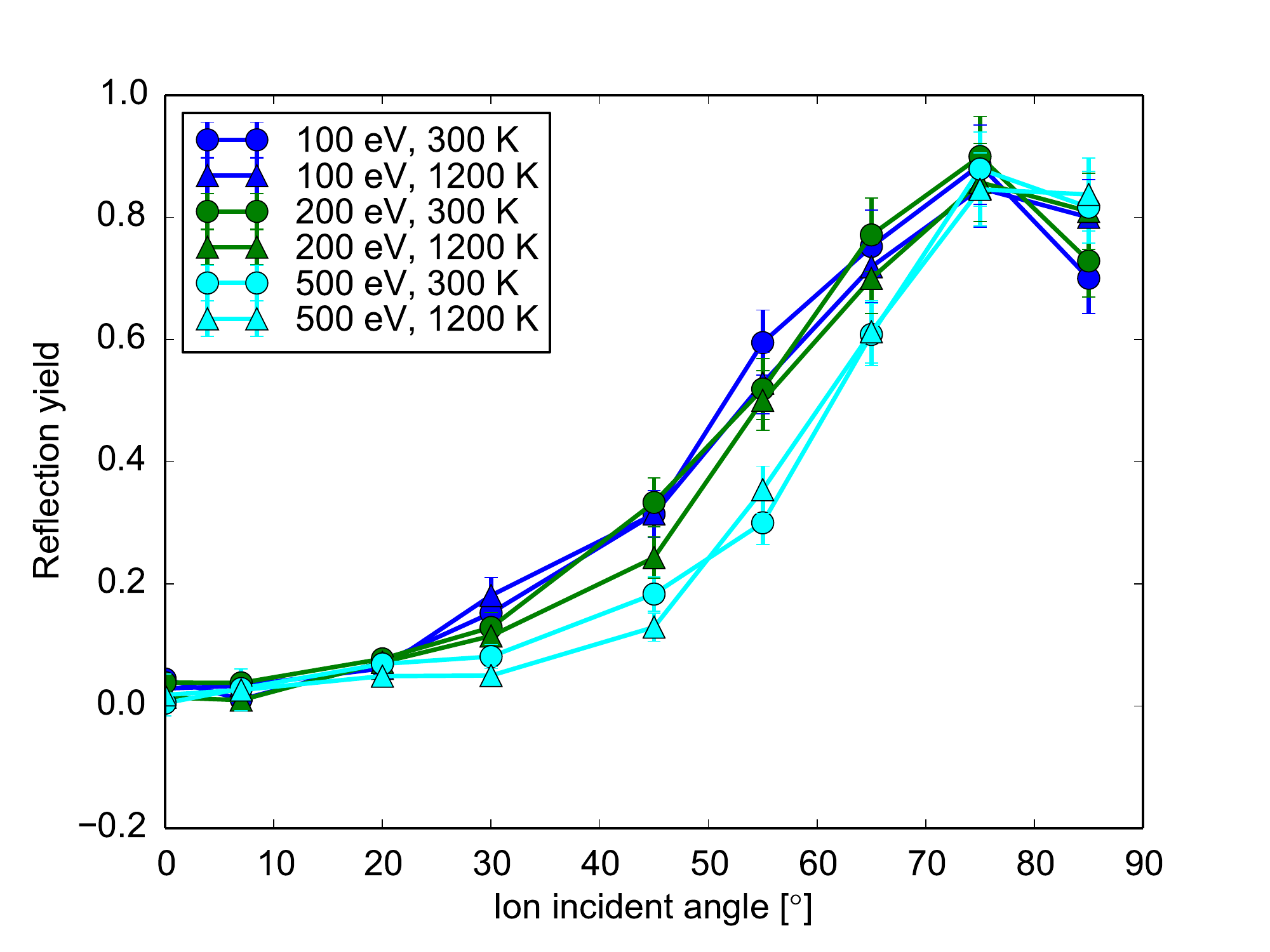}
\caption{Average reflection yield versus $\theta$ for different ion energies and temperatures.}
\label{fig:angle42} 
\end{figure}

We also analyzed the reflection yield for the ions impacting on the surface under inclined angles. These results are summarized in Fig.~\ref{fig:angle42}. The increase of the incident energy is somewhat reducing the reflection yield for the angles up to $\approx$ 60\textdegree, however, no significant dependence on the temperature is observed.  Changing the crystallographic orientations affected neither the sputtering nor the reflection yields.
 
\begin{figure}
\includegraphics[width=\columnwidth]{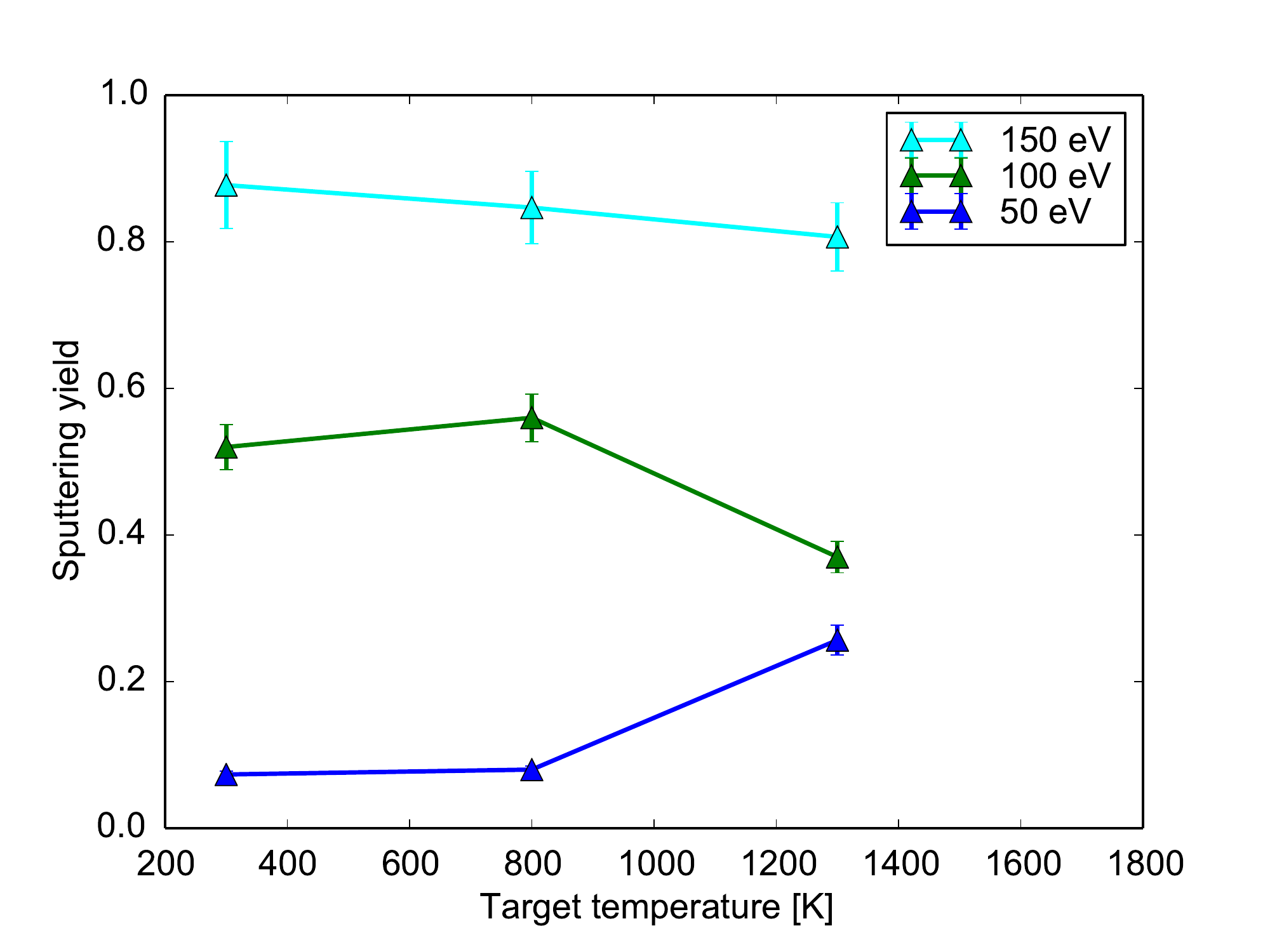}
\caption{Average sputtering yield versus target temperature for different projectile energies. The same incident angle $\theta = 5\degree$ was used in all cases.}
\label{fig:temperature_dependence}
\end{figure}

As the final analysis of temperature effect, we show in Fig. \ref{fig:temperature_dependence} the angular dependence of the sputtering yields for the ions of the plasma-relevant energies (50 eV, 100 eV and 150 eV \cite{Tim11b}) for different temperatures, 300--1300 K. The angle of incidence in this case was fixed slightly off-the-normal, $\theta=5\degree$, in all cases. Again, no dramatic change in the sputtering yield with temperature is seen and the sputtering yield for these incident energies remains below 1. This is in stark contrast with the results by Insepov et al., who predicted sputtering yields of $\sim$10 at 1300 K for the same ion energies \cite{Insepov2010AtomisticLinacs}.

\subsection{Effect of surface roughness}\label{sec:roughness}

\begin{figure*}
\centering
\subfigure[]{
\includegraphics[width=\columnwidth]{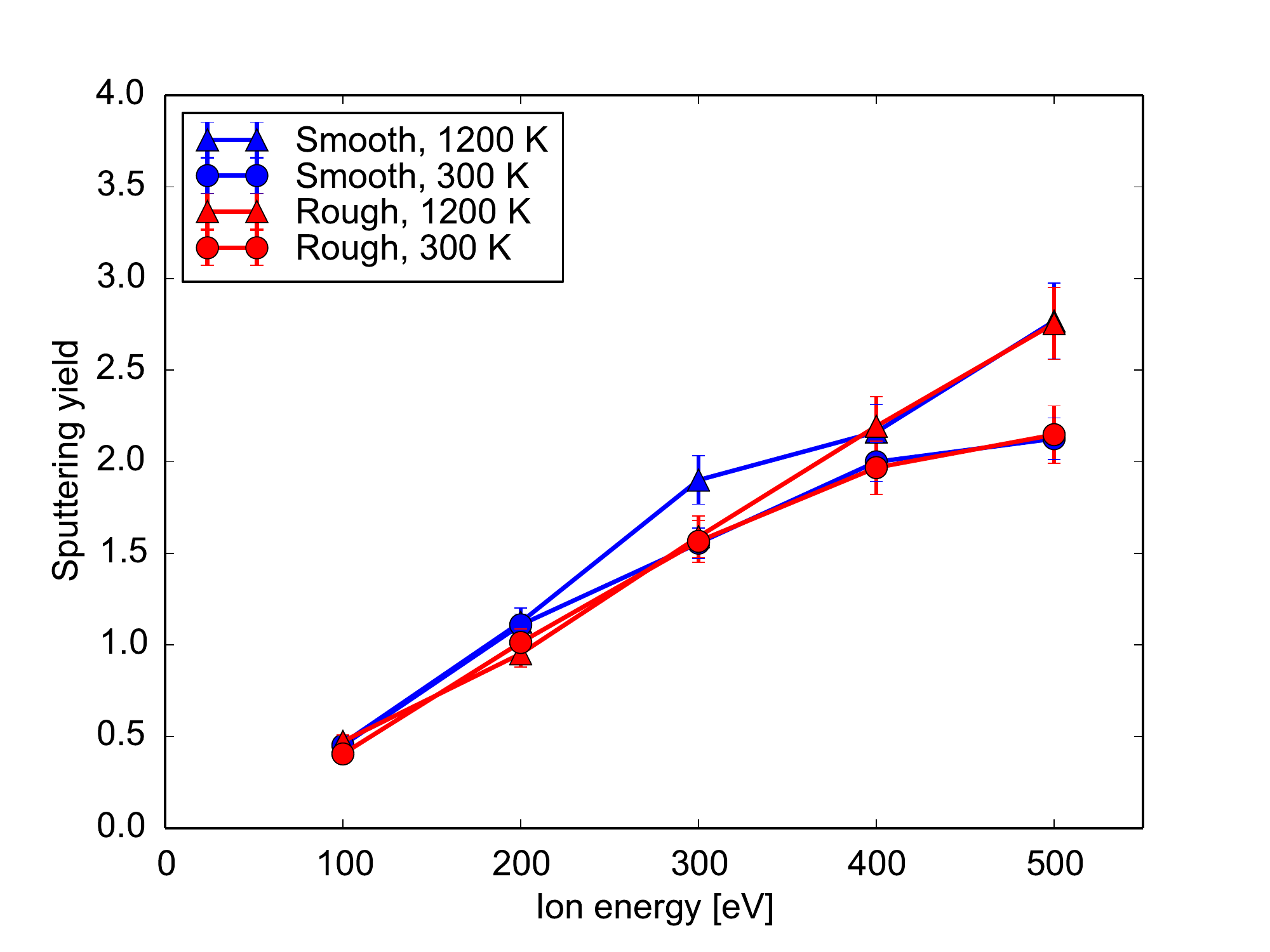}
\label{fig:rough1}
}
\subfigure[]{
\includegraphics[width=\columnwidth]{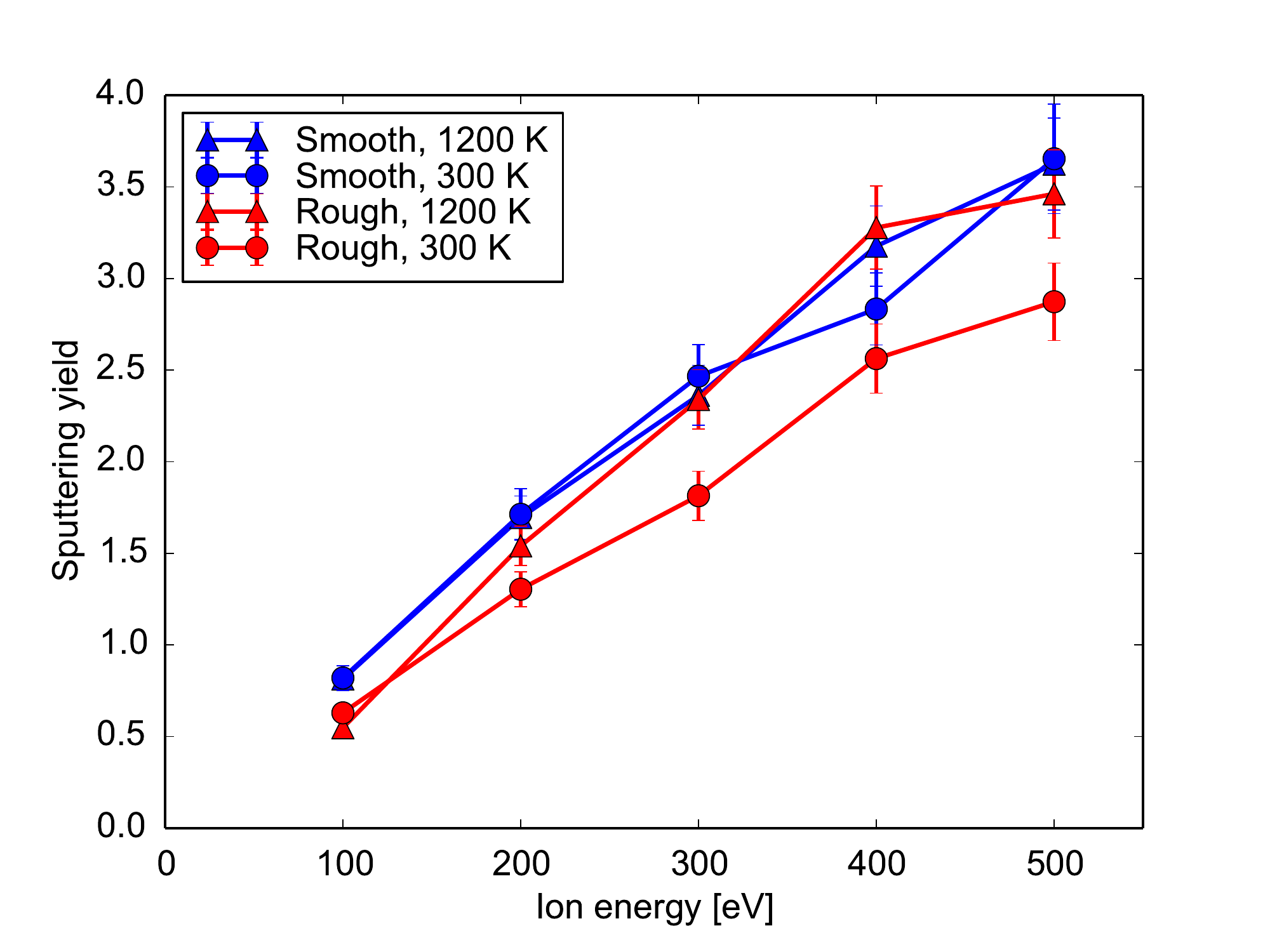}
\label{fig:rough2}
}
\caption{Averaged sputtering yield for smooth versus rough surfaces, using different incident angle distributions: (a) $\theta = 0\degree$, (b) $27\degree \leq \theta \leq 33\degree$. }
\label{fig:rough}
\end{figure*}

In Fig.~\ref{fig:rough}, we compare the energy dependences of the sputtering yields, $Y(E_k)$, from smooth and rough polycrystalline surfaces in the energy range between 100 and 500 eV and at two temperatures, 300 and 1200 K. $\overline{Y}(E_k)$ for the smooth surface was averaged over the three atomically smooth surfaces \{100\}, \{110\} and \{111\} (roughness $R \gg 1$ nm) and $\overline{Y}(E_k)$ for the rough  polycrystalline surface was averaged over the surfaces generated by KMC, see Fig. \ref{fig:KMC_surface} (roughness $R\approx1$ nm) with the same overall crystallographic orientation. 

Under the normal incidence (Fig.~\ref{fig:rough}a), the presence of a \{110\} surface decreases the sputtering yield due to intensive channeling in this direction. This effect becomes stronger with increasing incident energy in the considered range, increasing the difference in the sputtering yields. This is why, we observe the sputtering yields of the smooth polycrystalline surface gradually diverging  from the room temperature results with increase of the incident energy. The sputtering yields from the rough surface is surprisingly close in the values to the smooth surface. Apparently, the presence of small islands (see Fig.\ref{fig:KMC_surface}) does not affect the behavior of the surface under ion bombardment at least, in the considered range of the incident energies. 

We also see that the sputtering yields from the surfaces with low and high temperatures of the target differ more for smooth surfaces for energies above 200 eV, while the rough surfaces still result in closer values of the sputtering yields. Although the difference is not large, we believe that it can be explained by weaker channeling effect, since the channels may slightly close during the rough surface relaxation. However, some atomic cascades, which develop near the surface, cross the surface to a larger extent near rough features. This may lead to a somewhat higher sputtering yield, which may, in turn, accidentally coincide with the sputtering yield reduction resulting from the channeling effect. However, already the ions with the energy of 500 eV demonstrate that channeling is the key effect, which can explain the difference between sputtering yields from the surfaces with different temperatures, since both rough and smooth surfaces independently resulted in the same value of the sputtering yield. 

The tilted incidence shows clearer effect of surface roughness. These results are presented in Fig.~\ref{fig:rough}b. Here, the incident angle varied between $27\degree$ and $33\degree$, to take into account the spread of ion impact directions. 
In these simulations, we observe no effect of temperature on sputtering yield of the smooth surface (no channeling effect), however, the rough surface is sputtered more efficiently at higher temperature and higher incident energies. This is due to the presence of many atoms in the weaker-bonded positions (adatoms) on the rough features. 
However, overall the sputtering yield in this case is lower than that of the smooth surface, which can be explained by the effectively deeper position of cascades in case of the rough surfaces under the islands and hillocks on one hand and higher probability of redeposition of atoms sputtered with the low energies at small angles from the surface.

\section{Discussion and conclusions}\label{sec:DiscussionConclusions}

We studied the self-sputtering yield of Cu from collision cascades and heat spikes at different ambient temperatures by means of Molecular Dynamics simulations, in order to investigate how sputtering can be affected by the condition relevant to formation of plasma in an ultra high vacuum environment.

After performing almost 70,000 simulations of Cu self-sputtering at low incident energies ($< 500$ eV) with different initial conditions, such as crystallographic orientation, incident angle, surface roughness, and temperature, we concluded that the crystallographic orientation plays role only if the energy of incident ions is higher than 500 eV. Surprisingly, we saw a good comparison of tendency in angular dependence between our MD results and SRIM calculations down to very low energy (100 eV), while Yamamura's semi-empirical formula gave a good agreement for the highest studied energies.  

We did not observe any temperature effect on angular dependencies of sputtering and reflection yields. Overall, only very high temperature simulations, near the melting point of Cu, demonstrated a somewhat observable dependence on the  temperature. We related this increase to reduction of channeling effect at 1200 K, since the simulations with the off-normal incidence did not show any difference even at very high temperatures.

Rough surfaces with features $\sim$1 nm high were found to sputter less efficiently and rather independent of the temperature at the normal incidence, while inclined incidence promoted a stronger sputtering with temperature increase. 

  Overall, the simulations carried out in this work show that neither sample
temperature, incidence angle, nor surface roughness enhances the
direct sputtering from individual collision cascades enough to explain how a plasma
can form in ultra-high-vacuum electric breakdowns. This indicates
that additional atom sputtering or evaporation effects must be present
under the conditions when a vacuum arc plasma builds up. These could
possibly be multiple overlapping collision cascades leading to a nonlinear
sputtering enhancement \cite{Tim09a}, or thermally enhanced adatom
evaporation that may enhance atom desorption between the actual collision
cascades \cite{Sch06b}. A further possibility is additional
surface heating induced by the field emission electron current.
Further studies of the effect of ion flux in synergy with electronic heating effects should
be pursued to understand the process of growing plasma densities.

In conclusion, we find no major anomalous increase of sputtering yield in single ion impacts by varying crystallographic orientations, incident angles or surface roughness. Neither would a sudden local increase of the target temperature be able to explain plasma build-up in RF cavities.

\section*{Acknowledgement}

This work was supported by Academy of Finland (Grants No.\;285382 and No.\;269696), Estonian Research Council (Grants PUT 57 and PUT 1372) and MEPhI Academic Excellence Project (Contract No. 02.a03.21.0005). Computing resources were provided by the Finnish IT Center for Science (CSC) and High Performance Computing Center of University of Tartu.


\bibliography{metspalu,pub/vjansson_publications,pub/vjansson,general,clic_fd}

\end{document}